\documentclass[sigconf]{acmart}

\setcopyright{none}
\renewcommand\footnotetextcopyrightpermission[1]{}
\setcopyright{none}
\makeatletter
\renewcommand\@formatdoi[1]{\ignorespaces}
\makeatother

\usepackage{subfig}
\usepackage{xspace}
\usepackage{xcolor}

\newcommand{\one}{({\em i}\/)\xspace}
\newcommand{\two}{({\em ii}\/)\xspace}
\newcommand{\three}{({\em iii}\/)\xspace}
\newcommand{\four}{({\em iv}\/)\xspace}
\newcommand{\five}{({\em v}\/)\xspace}

\def\eg{\emph{e.g.,}\xspace}
\def\etc{\emph{etc.}\xspace}
\def\ie{\emph{i.e.,}\xspace}
\def\etal{\emph{et al.}\xspace}
\def\vs{\emph{vs.}\xspace}

\def\wa{{WhatsApp}\xspace}

\begin{document}

\title{Jettisoning Junk Messaging in the Era of End-to-End Encryption: A Case Study of WhatsApp}

\author{Pushkal Agarwal}
\email{pushkal.agarwal@kcl.ac.uk}
\affiliation{%
  \institution{King's College London}
  \country{United Kingdom}
}

\author{Aravindh Raman}
\email{aravindh.raman@telefonica.com}
\affiliation{%
 \institution{Telefonica Research}
 \country{Spain}
}

\author{Damiola	Ibosiola}
\email{d.i.ibosiola@qmul.ac.uk	}
\affiliation{%
  \institution{Queen Mary University of London}
  \country{United Kingdom}
}

\author{Nishanth Sastry}
\email{n.sastry@surrey.ac.uk	}
\affiliation{%
  \institution{University of Surrey}
  \country{United Kingdom}
}

\author{Gareth Tyson}
\email{gtyson@ust.hk}
\affiliation{%
  \institution{Hong Kong University of Science \& Technology, Hong Kong}
}

\author{Kiran Garimella}
\email{kiran.garimella@rutgers.edu}
\affiliation{%
  \institution{Rutgers University}
  \country{United States of America}
}

\begin{abstract}

\wa is a popular messaging app used by over a billion users around the globe. 
Due to this popularity, understanding misbehavior on \wa is an important issue.
The sending of unwanted junk messages by unknown contacts via \wa remains understudied by researchers, in part because of the end-to-end encryption offered by the platform.
We address this gap by studying junk messaging on a multilingual dataset of 2.6M messages sent to 5K public \wa groups in India.
We characterise both junk content and senders. We find that nearly 1 in 10 messages is unwanted content sent by junk senders, and a number of unique strategies are employed to reflect challenges faced on \wa, \eg the need to change phone numbers regularly.
We finally experiment with on-device classification to automate the detection of junk, whilst respecting end-to-end encryption. 

\end{abstract}



\maketitle

\renewcommand{\thefootnote}{}
\footnote{\texorpdfstring{$^\star$}{}A PREPRINT OF ACCEPTED PUBLICATION AT The Web Conference (WWW) 2022.}
\renewcommand{\thefootnote}{\arabic{footnote}}

\lhead{Jettisoning Junk Messaging in the Era of End-to-End Encryption: A Case Study of WhatsApp}
\rhead{Agarwal P., et al.}

\section{Introduction}

WhatsApp is the most popular messaging app in the world, with over 1.5 billion active users each day. 
With this massive popularity, \emph{unwanted messages} (such as spam) have become a challenge for \wa.
However, unlike existing messaging systems (\eg email, Twitter), where platforms can read the content, \wa follows an end-to-end encryption model where the content is not accessible.
Whereas this offers stronger guarantees on privacy, it makes content moderation difficult.

Although \wa has made progress \cite{jones2017whatsapp} in detecting users who send unsolicited messages to individuals, there is no solution for users who send such messages to \textit{public WhatsApp groups} \cite{garimella2018whatsapp}. These are groups where the admins publicly share a link (\eg via a website or social platform) to join the group. 
\wa provides strong protection from being contacted by strangers, allowing users to easily block unsolicited messages from those not in their contact list. In contrast, as long as a user is a member of a public \wa group, they cannot avoid messages that strangers may send, regardless of whether the messages are germane to the group or not. Thus, these strangers can abuse public \wa groups by sending messages that are irrelevant to the purpose of a group, \eg links to adult content or sexual services, phony job offers \etc

Understanding the nature of unwanted junk messages in public groups is therefore vital for improving users experience on \wa and other messaging platforms. 
For instance, receiving excessive notifications can be highly disruptive and undermine the responsiveness to legitimate notifications~\cite{pielot2014situ,  mehrotra2016my}.
In this paper, we use the term ``junk messages'' to capture a range of messages that are unlikely to be of interest to the general group membership. This may be because the content is unsavoury (\eg links to adult content in political groups) or simply irrelevant (\eg asking people to fill out surveys). We therefore define junk messages as: ``\emph{those which are not considered of interest or suitable by admins for a group, leading such posters to be removed}''.

To understand the nature of such misbehaviour we focus on India, a country where over 400 of the 460 million people online are on \wa.\footnote{https://www.cjr.org/tow\_center/india-whatsapp-analysis-election-security.php}
Given the population we are studying (mostly first time Internet users from India), the susceptibility of believing the content in junk message (\eg fake job offers) is high.
As a common topic that attracts interest from across the population (all languages and states), we focus on \emph{national politics}, and gather 2.6 million messages from 5,051 public political \wa groups in India (\S\ref{sec:data}) which are posted before, during and after the elections in 2019.

In this paper, we analyse the characteristics of junk messages (\S\ref{sec:whatisspam}) as well as the actions of junk senders (\S\ref{sec:spammers}). While we do find parallels with junk posted in other platforms, we see distinct trends which are unique to \wa.
For example, as with platforms like Facebook~\cite{redmiles2018examining},
\wa junk message consists of topics such as job advertisements and click bait often embedded within URLs.
We also observe unique patterns to \wa, \eg junk senders using multiple phone numbers to spread the same messages over a small number of days (note that phone numbers are typically more difficult to obtain than email addresses in India). 
These unusual traits lead us to explore junk senders' evasion tactics. We see many joining and leaving groups multiple times, thereby avoiding being removed by admins, and nearly a quarter of all phone number changes are performed by junk senders. 
Finally, we investigate ways to support admins  (\S\ref{sec:modeling}) by devising end-to-end encryption compliant classifiers to identify junk messages. We obtain good performance, with accuracy scores exceeding 0.85. This offers a foundation for future moderation efforts in the era of end-to-end encrypted messaging.

\section{Dataset \& Methodology}
\label{sec:data}

\subsection{Data collection methodology}

We take inspiration from prior work, which found that public groups that discuss politics are widely used in India
\cite{lokniti2018, saha2021short} and Brazil \cite{reuters2019report,reis2020can}. 
Our study is based on data from \textit{public} \wa groups discussing politics in India. 
We select this focused subset of groups, as it allows us to more easily identify out-of-scope messages (\ie junk), as these pertain to topics outside of politics. 
To this end, we searched for \wa group links (\texttt{chat.whatsapp.com}) on Facebook, Twitter and Google during November 2018 using an extensive set of 349 manually curated keywords\footnote{List available: \url{http://tiny.cc/curated-keywords}} from multiple Indian languages (including English)
relating to politics in multiple languages.
This yielded 5,051 groups.
These are typically created by political parties or party supporters in order to reach an audience which is only available via \wa.
Hence, most of these groups have a well defined organizational structure~\cite{banaji2019whatsapp}.
Note, due to end-to-end encryption, \wa is limited in its ability to moderate content in these groups. Instead, moderation is mostly left to group admins, who also have powers to remove users.


With this list of 5,051 groups, we use the toolkit from~\cite{garimella2018whatsapp} to collect data. 
We collect all messages between October 2018 and August 2019.
Across the 5,051 groups, we collect 2.6 million messages posted by over 172K unique users over a period of 302 days.
We also record 437K action events, covering actions taken by users including entering or leaving groups and changing phone numbers. 
Table~\ref{table:actions} summarizes the actions performed by users within a group.

\noindent\textbf{Ethics note}: Our data collection abides by the terms of service of \wa and was considered exempt by the Institutional Review Board at MIT. All data was anonymised before analysis, and any personally identifiable information was masked. All phone numbers were one-way hashed, after extracting the country code.

\begin{table}
    \centering
    \caption{Actions captured within a group.}
    \vspace{-12pt}
    \resizebox{\columnwidth}{!}{
    \begin{tabular}{p{2.3cm}|p{3.4cm}|p{0.99cm}|p{0.85cm}} 
         \hline
         \textbf{Action} & \textbf{Description} & \textbf{Action Counts} & \textbf{Unique Users}\\ 
         \hline
         \emph{added} & added by a member  &61k& 37k \\ 
         \hline
         \emph{added\_by\_admin} & added by a group admin & 73k&49k \\ 
         \hline
         \emph{joined\_via\_link} & joined via an invite link& 132k&54k \\ 
         \hline
         \emph{left} & left the group & 154k&73k\\ 
         \hline
         \emph{removed} & removed from a group& 9k&7.3k \\ 
         \hline
         \emph{number\_changed} & changed from one number to another & 6k&1.5k\\ 
     \hline
    \end{tabular}
    }
    \label{table:actions}
    \vspace{-\baselineskip}
\end{table}

\subsection{Message pre-processing}
\label{sec:preprocessing}
Our dataset consists of 2.6M messages out of which only 1.4M are unique. We filter and cluster these messages to have a unique and distinct set of messages. 

\subsubsection{Message filtering.}
Because the content of the message is important in identifying junk, we focus on the four top languages (Hindi, English, Telugu and Tamil) which collectively represent 74\% of messages in our dataset.\footnote{
We tag messages with their language using Google's Language Detector~\cite{ooms2018cld2} in R.}
We also remove messages containing just URLs (no accompanying text), boilerplate content such as `hi' or `good morning', and, messages consisting solely of emojis, which constitute around 25\% of our data. 
We filter such content to avoid characterising low entropy posts -- such messages could fall under both on-topic and off-topic as junk~\cite{gupta2019good}. As later discussed, the percentage of boilerplate messages among junk posters is 7\% (12\% for legitimate users). 
Filtering these out, we are left with 766K messages from 2.6 million messages which is in turn quicker to cluster as we discuss in next section.

\subsubsection{Message clustering.}
We qualitatively observe that many messages are close variants of each other.
To group together near-similar variants of the same message, we use MinHash and Locality Sensitive Hashing (LSH)~\cite{gionis1999similarity,mullen2015textreuse}.
We find that the best clustering performance is obtained by using 10 min-hashes in 5 bands.
We made this choice by taking 100 near-identical messages and 100 distinct messages, and experimenting with a range of parameters to derive the optimal for separating out these two sets.

Overall, this results in 73K clusters, with an average of 10 (median 4) messages in each cluster. 
As a quality check, we randomly select 100 random clusters from each of the 4 languages and manually verify that 97\% are similar (the rest are from a bot service~\cite{duta_bot}). 
Of the 73K clusters, 25.4K have at least 5 messages, and contain 420K messages in total. These 420K messages, which we term  ``frequently sent messages'' (sometimes shortened to ``frequent messages'') are the object of study for the rest of the paper, except when studying URLs (\S\ref{sec:urls-phones}), where we include all the messages containing URLs.

\subsection{Junk and non-junk annotation}
\label{sec:annot}

We next describe annotation of messages as `junk' or `non-junk'. 

\subsubsection{Manual annotation.}
We start by identifying a seed set of users who were manually removed from at least two groups by their admins.
Recall that according to our earlier definition, messages that trigger users to be removed, are considered junk candidates. We therefore conjecture that users who are manually removed by different group administrators more than twice (257 users) may provide a useful set of candidate junk messages. We later confirm that this is true: over 90\% of messages sent by these users are later annotated as junk and are often accompanied by comments by admins such as ``pls don't spam in this grp with such posts''.

We first extract the 68K messages (grouped into 1,004 clusters) sent by these 257 users and manually annotate each cluster as `junk' or `non-junk'.
We perform the annotations in two stages.
First, two English-speaking `lead' annotators (domain experts) work independently to label the 220 English clusters contained within the 1,004 clusters. 
Since a large portion of junk content is self-evident, the inter-rater agreement (Cohen’s-Kappa) is high (0.96). The remaining disagreements were discussed to reach full agreement. Based on the labeling of 220 clusters, two lead annotators prepare guidelines for further annotation on larger datasets by multiple members of the annotation team.

Following this initial step, which validated the ability to identify junk messages with high agreement levels, we progress to the second stage. We recruit 2 annotators for English, 3 for Hindi and 1 each for Telugu and Tamil.
We ask them to further categorise the junk messages into the following categories:
\one~Promotion messages (if non-political promotion as these groups are for political discussions).
\two~Adult content related messages.
\three~Invitations to register for external services via links, often for money.
\four~Offers to earn money, win prizes etc.
Or \five~Anything which looks `suspicious' and not relevant to the group at large. 
In cases where annotators find something suspicious but hard to classify, they were allowed to consult the lead annotators who prepared the guidelines. 
With this guidance, the set of 784 clusters were given to annotators (native speakers of Hindi, Telugu and Tamil).
Each annotator received the subset of messages in their native language, and were given identical guidelines (as described above).
In total, the above two steps result in 63K messages (from 663 clusters)  being tagged as junk and 5K messages (341 clusters) being tagged as non-junk.  \looseness=-1

\subsubsection{Semi-Automatic Annotation.}
The above produces a ``gold standard'' set of annotated messages.
We next complement this with a semi-automatic annotation process to broaden our analysis.
Here, we construct a dictionary of words that are used at least 5 times in the 63K messages classified as junk above.
We then manually clean the list to obtain a set of 324 high precision junk words.\footnote{The threshold 5 was set by manually inspecting the results obtained by selecting various values from 3 to 10 in order to obtain a list of junk words in all four languages. Stop words were removed in all languages using multi-lingual dictionary from `stopwords' library in R.) Some example words include free, job, iphone, samsung, IELTS, sex, click, sales and vacancy.}
We then search for the occurrence of these junk words in the entire database of 420K frequently sent messages in Hindi, English, Telugu and Tamil.
In each cluster where we find one or more of the junk words, the annotators examine the cluster and validate it as `junk' or `non-junk' using the same approach as above. Finally, we obtain a labelled dataset containing 295K junk (from 3.5K clusters) and 112K non-junk (from 3.2K clusters) messages as summarised in  Table~\ref{tab:annotation}.\looseness=-1

\subsubsection{Validation.}
We note that the above annotations are inherently subjective. 
Hence, to give greater confidence that our concept of junk and non-junk is reasonable, we finally create an independent panel of 11 assessors (distinct from the annotators) who look at a randomly chosen set of 100 messages (split into 50 junk and 50 non-junk messages). All assessors were Indian, and collectively represent 7 states with different age groups and genders. The panel were only told that these messages are from public \wa groups related to Indian politics and were asked to mark whether they think such messages are `junk' or `non-junk' based on their own personal understanding. We find that there is a median agreement of 95\% across the labels for all the independent assessors (labelling 100 messages without any guidelines) and our annotators (labelling large scale data with the prepared guidelines).

\begin{table}[t]
    \caption{Summary of the annotations set.}
    \vspace{-12pt}
    \label{tab:annotation}
    \centering
    \resizebox{0.9\columnwidth}{!}{
    \begin{tabular}{l|l|l|l}
         \hline
         \begin{tabular}[c]{@{}l@{}}\textbf{Message}\\\textbf{type}\end{tabular}&
         \textbf{From}&
         \begin{tabular}[c]{@{}l@{}}\textbf{Unique}\\\textbf{messages}\end{tabular} &
        \begin{tabular}[c]{@{}l@{}}\textbf{Total}\\\textbf{messages}\end{tabular} \\
         \hline
         \hline
         Junk & Removed users& 663   & 63K \\
          Non-Junk & Removed users & 341  &   5K  \\
          Junk & Semi-automation & 2.8K  & 232K \\
          Non-Junk & Semi-automation  & 2.9K      &107K     \\
         \hline
    \end{tabular}
    }
    \vspace{-\baselineskip}
\end{table}

\begin{figure}[ht]
\centering
 \includegraphics[width=0.85\columnwidth]{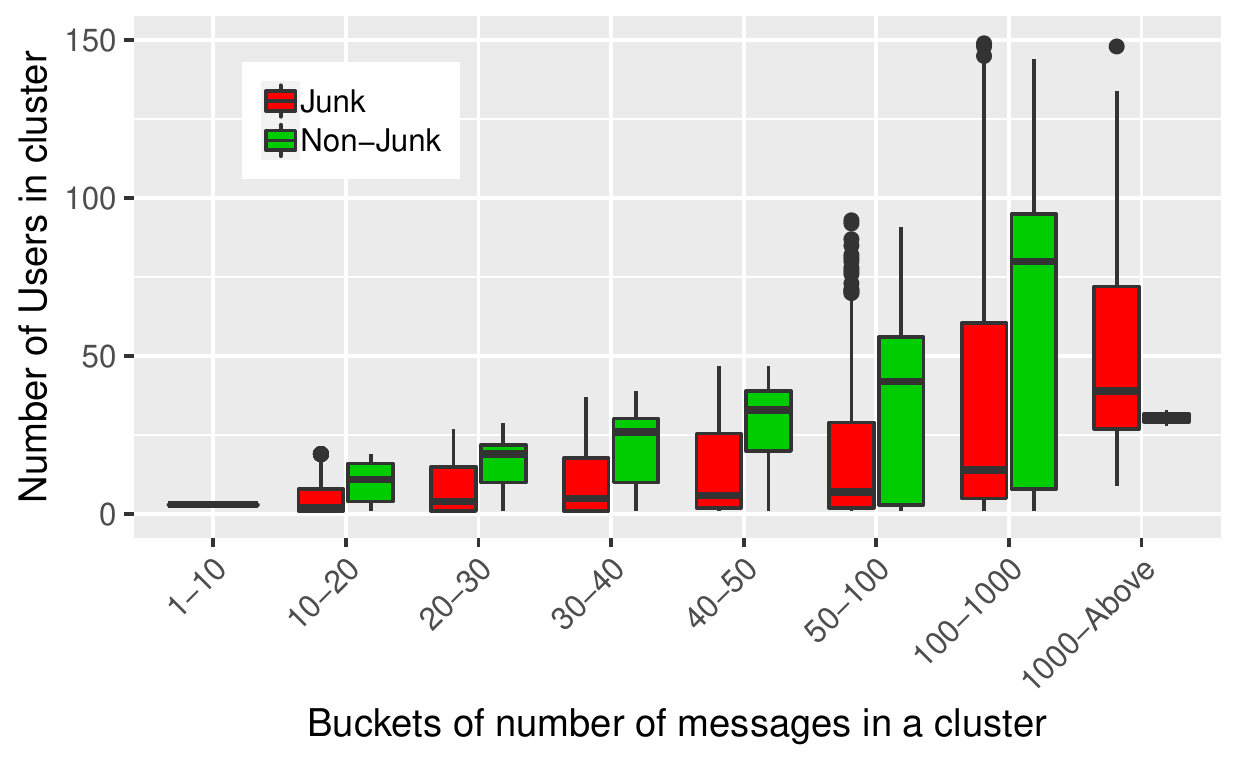}
 \caption{Numbers of users spreading a message, indexed by number of times the message or its close variants are seen.}
\label{fig:userMessageCluster}
\vspace{-\baselineskip}
\end{figure}


\section{Characteristics of Junk Content}
\label{sec:whatisspam}


We first look the nature and content of \wa junk \emph{messages}.

\subsection{Understanding the scale of junk}
\label{sec:scale}

\begin{figure*}
\hspace*{2.5cm}\includegraphics[width=.37\linewidth]{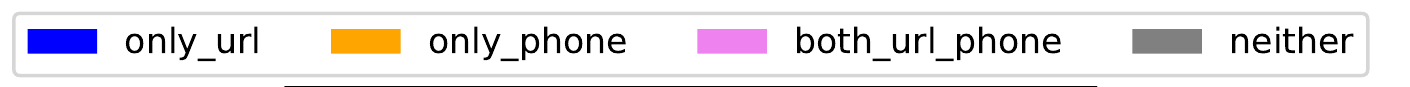} 

\subfloat[\label{fig:category}]{\includegraphics[width=.37\linewidth]{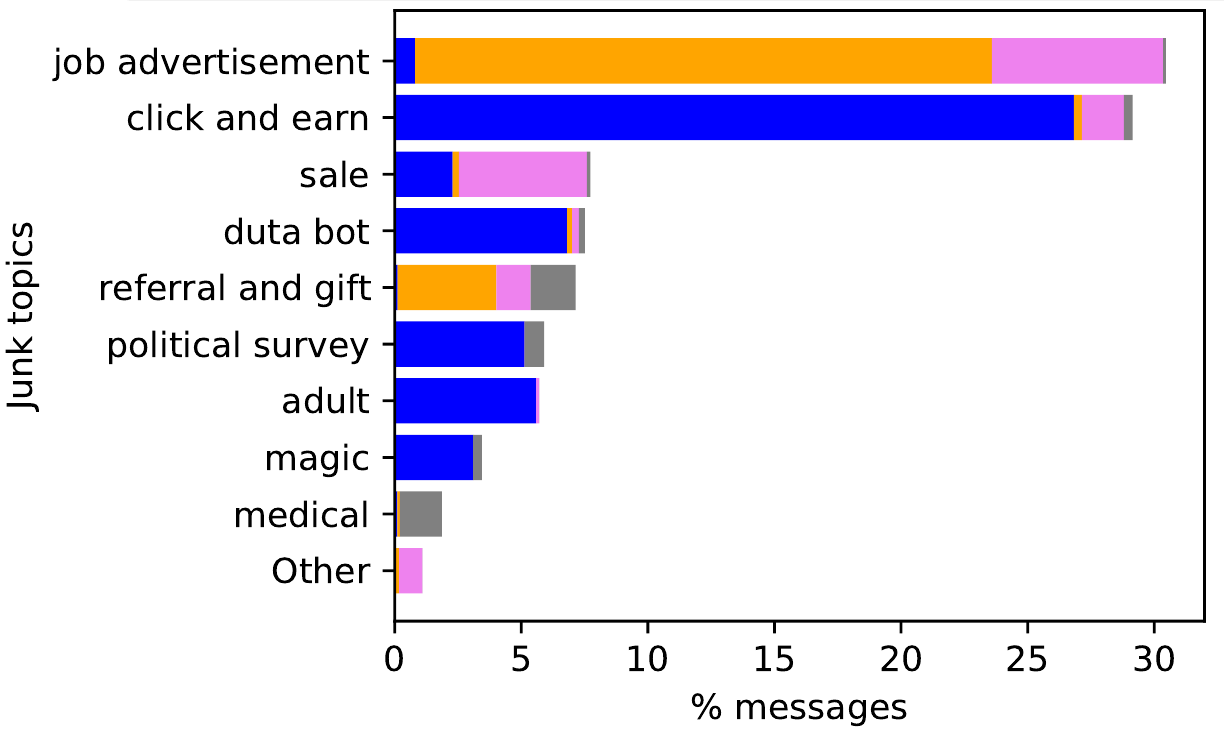}}
\subfloat[\label{fig:urlphonecomp}]{\includegraphics[width=.178\linewidth]{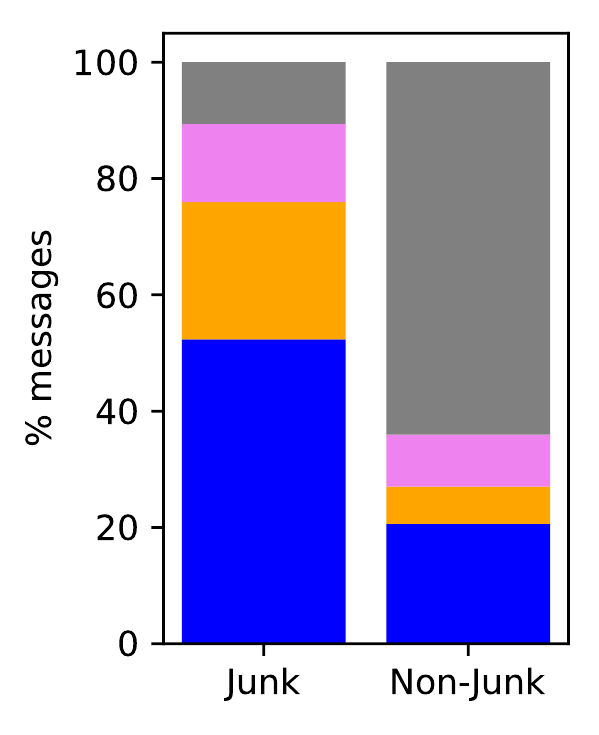}}
\subfloat[\label{fig:urlcat}]{\includegraphics[width=.4\linewidth]{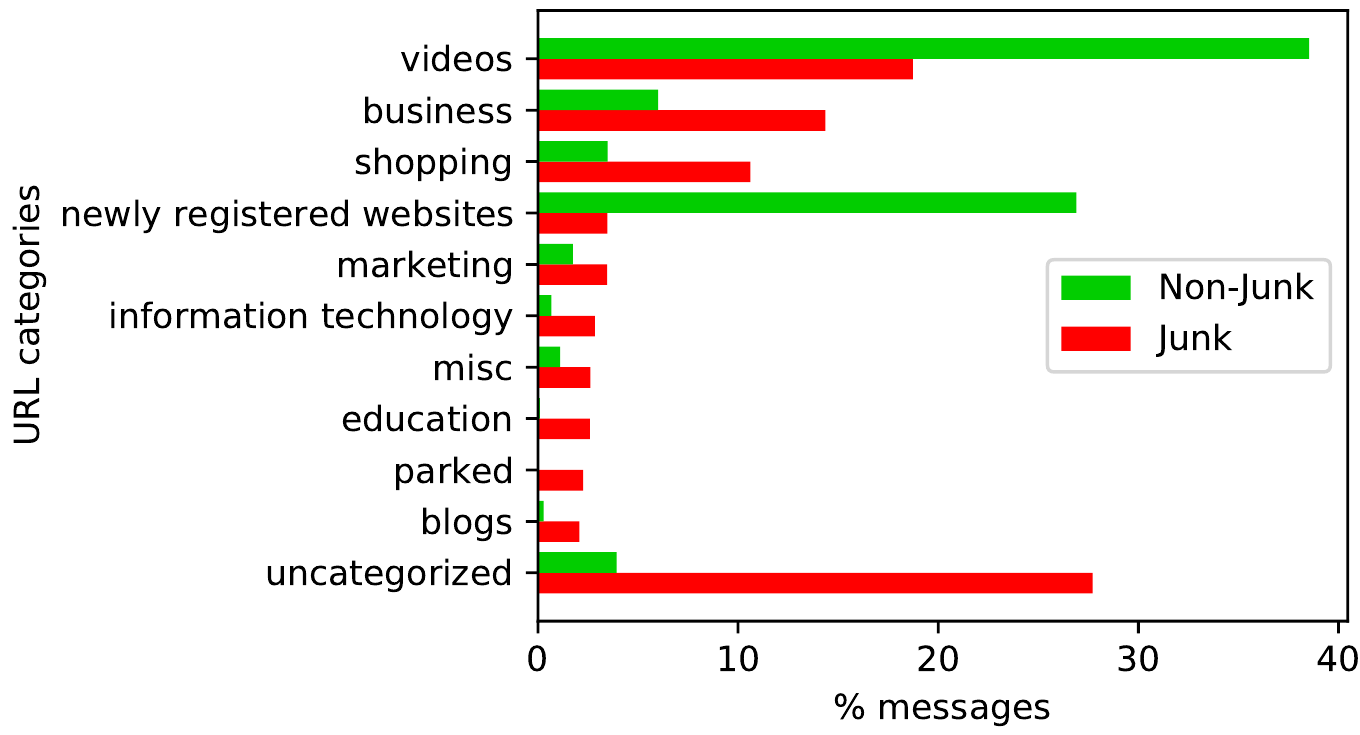}}

\caption{(a) Topics found in the top 100 clusters of Junk (48\% of all Junk messages); (b) Overall fraction of URLs and phone numbers found in the content of messages; (c) Categories of URLs as identified by VirusTotal in messages.}
\vspace{-0.5cm}
\end{figure*}

We first consider two aspects of scale: the number of times a message is posted, and the number of users who send it. Each message cluster obtained (see \S\ref{sec:preprocessing}) may contain several messages that are closely related variants. Figure \ref{fig:userMessageCluster} groups message clusters into different buckets based on the number of times those messages are found in our data, across different groups.

We find that clusters containing junk messages have larger numbers of messages  on average, although the median numbers are similar (24 for junk clusters \vs 23 for non-junk clusters; compared to mean 83.6 in junk \vs 35 for non-junk).  
This indicates a highly skewed distribution of messages per cluster, particularly for junk clusters. For instance, there are over 37 clusters of junk with more than 1000 messages, the largest cluster having over 27K messages. In contrast, only 2 non-junk clusters have more than 1000 messages and the largest having 2,065 messages.
These are videos URLs and associated text related to political speeches.

We also see that popular clusters, which contain messages that are forwarded many times, inevitably involve more users. However, Figure \ref{fig:userMessageCluster}  
shows that for messages in junk and non-junk clusters of similar sizes (\ie in the same ``bucket'' size on the x-axis), the junk clusters are driven by fewer users. This indicates that junk is disseminated in a more proactive manner by a smaller set of users.

\subsection{Understanding junk content}

\subsubsection{Junk topics.}
\label{sec:spamtopics}
To explore the topics discussed within junk, we employ three annotators to manually examine messages in the top 250 clusters of junk messages.
By performing an initial qualitative analysis, we identify 10 core topics. We then ask the annotators to categorise the messages from the top 250 clusters into these 10 topics. 
The top 250 clusters comprise 61\% (181K messages) of the total number of junk messages (295K) in our dataset. Each message cluster is examined by at least one annotator and the category label applied was checked by a second annotator.  All differences of opinion were resolved in a discussion and we finally obtained 100\% inter-agreement between annotators.

Figure \ref{fig:category} captures the relative frequencies of the topics, as well as the frequency of URLs and phone numbers within the messages.
We emphasise that this taxonomy is based on the fact that these messages are post to \emph{political} groups. Some of these may not be considered junk if posted to other types of groups.
\\
\textbf{Job Advertisements.} Comprising nearly 30\% of our annotated junk, the most widespread junk is advertisements for jobs. Nearly all (97\%) of job advertisements provide a contact phone number, or a phone number as well as a URL. Of course, these may be genuine job advertisements, although they are off-topic junk for political \wa groups. 
Though we are not able to identify if these were genuine job ads, the template structure of these ads makes us believe (since such users are removed) these are junk and could involve a scam. Interestingly, over 95\% of junk in Telugu language forums consist of job advertisements.
    \\
    \textbf{Click and Earn.} These comprise 29\% of junk messages, and ask users to click on a URL, promising a reward. 98\% of these messages contain a URL, but no phone number. 
    \\
    \textbf{Sales.} These constitute 7.7\% of junk and offer items for sale. 65\% of these messages contain URLs and a phone number, and could be genuine items for sale. 
    \\
    \textbf{Referral and gifts.} These junk messages (7.5\%) offer a gift in return for referrals of users to an online service subscription, and consist mostly of a URL to click.
    \\
    \textbf{Duta Bot.} These are (benign) junk messages (7.1\%) sent by a news bot service called Duta Bot~\cite{duta_bot} comprising regular news or sports updates. Note that many admins do not welcome the Duta Bot posts: they are regularly removed from groups. We find that, 26 bots who posted 1000 junk messages are removed by admins.
    \\
    \textbf{Adult.} These (5.7\%) mostly contain URLs that lead to adult websites or offer adult sex-related services. 
    \\
    \textbf{Political Survey}. These (5.6\%) mostly contain junk URLs that invite users to participate in fake political surveys.
    \\
    \textbf{Magic.} These (3.4\%) messages contain text which asks user to forward a message to experience something supernatural, \eg ``Forward and see magic: your phone battery will get charged to 100\%''.
    \\
    \textbf{Medical.} These (1\%) messages offer treatment for common and often embarrassing ailments, \eg ``ayurvedic treatment for piles''.
    \\
    \textbf{Other.} Approximately 2\% cannot be categorised into any of the above groups and consist of junk such as ``daily event update''.

\subsubsection{URLs and Phone Numbers.}
\label{sec:urls-phones}

As shown above, a significant number of junk messages contain URLs (167K messages) and phone numbers (74K messages).  
Figure \ref{fig:urlphonecomp} compares the fraction of junk and non-junk messages that contain phone numbers, URLs or both. 
We see a marked difference between junk and non-junk, with nearly 90\% of junk messages containing either a phone number, a URL or both (in contrast to just 36\% for non-junk). We also notice 19.4K unique phone numbers present in the \textit{content} of the messages (note that for ethical reasons, all numbers are one-way hashed before analysis). Out of this, only 9.5\% of the numbers were found to be senders of any messages, and 85\% of these were junk. 

We next take the 56.8\% of junk and 27.5\% of non-junk messages that contain URLs. To explore the nature of these URLs, we use VirusTotal\footnote{\url{https://www.virustotal.com/}} to classify each according to its type of activity~\cite{ikram2019chain,kim2015detecting}.
For 81.1\% of domains we identify the category associated with the domain. Figure~\ref{fig:urlcat} shows the distribution of URL categories for both junk and non-junk.
We see that video URLs are popular in junk messages, as well as newly registered sites (manual analysis reveals these are primarily news websites, \eg \texttt{upchaupal.com}). 
In contrast, junk messages carry far more business and shopping URLs. 
Worryingly, 1\% of the messages carry URLs marked as `elevated exposure', \eg \texttt{apkmaster.xyz}.
These are sites that camouflage their true nature or identity, or that include elements suggesting latent malign intent.\footnote{\url{https://bit.ly/34aBZ3V}}
Note, a notable fraction of junk emails are classed as ``uncategorized''. This is because these junk messages often contain a unknown fringe URLs, that are not present in the domain classifier service.

To explore these risks, we run all URLs through the 73 antivirus engines in VirusTotal. Overall, 26\% of URLs in junk messages are tagged as dangerous by 3 or more antivirus, and 15\% are tagged by 9 or more \eg \texttt{trycryptocoins.com}, \texttt{amazon.bigest-sale-live.in}.
In contrast, only 1.5\% of non-junk were tagged as malicious by 3 or more engines (0.42\% by 9+ engines).


\section{Junk senders and their Actions}
\label{sec:spammers}

Next we look at the \emph{users} who produce junk in terms of their locations, temporal patterns and group membership.

\subsection{Operational definition of junk senders}
\label{sec:opdefspammer}

\begin{figure}[t]
    \centering
    \includegraphics[width=0.85\columnwidth]{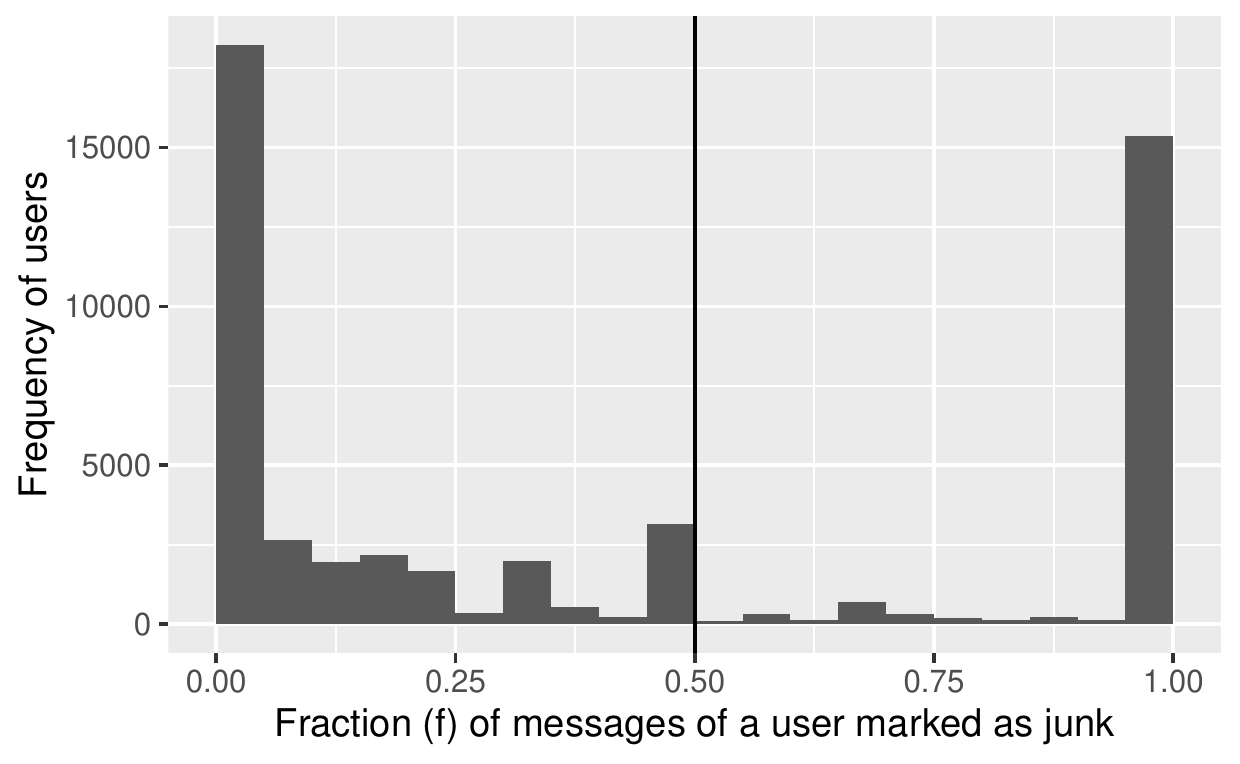}
    \caption{Fraction $(f)$ of messages of a user marked as junk. A user is considered a junk sender if they posted more than $f=50\%$ junk messages.}
    \label{fig:spammercdf}
    \vspace{-\baselineskip}
\end{figure}

Our methodology identifies junk messages by their content rather than junk senders directly. Some junk messages may be inadvertently posted by enthusiastic or na\"ive users who do not realise it is junk. 
Figure \ref{fig:spammercdf} plots the fraction, $f$, of messages by a user that are marked as junk with respect to total messages. As expected, this follows a bimodal pattern, with junk senders on one end (nearly 100\% of their messages are junk) and non-junk senders at the other end (with almost no junk messages). This suggests that users with a junk fraction beyond any reasonable threshold will capture all intentional spammers.  
In this section, we adopt an operational definition of junk senders as any user who has posted more than $f=50\%$ messages that our methodology identifies as junk (our results are robust to other similar thresholds). 
Using this methodology, we identify 17.6K users as junk senders and 32.9K as non-junk senders who share a total of 239K and 1.3M messages, respectively.

\subsubsection{Distribution of junk senders across groups}

\begin{figure}[t]
    \centering
    \includegraphics[width=0.85\columnwidth]{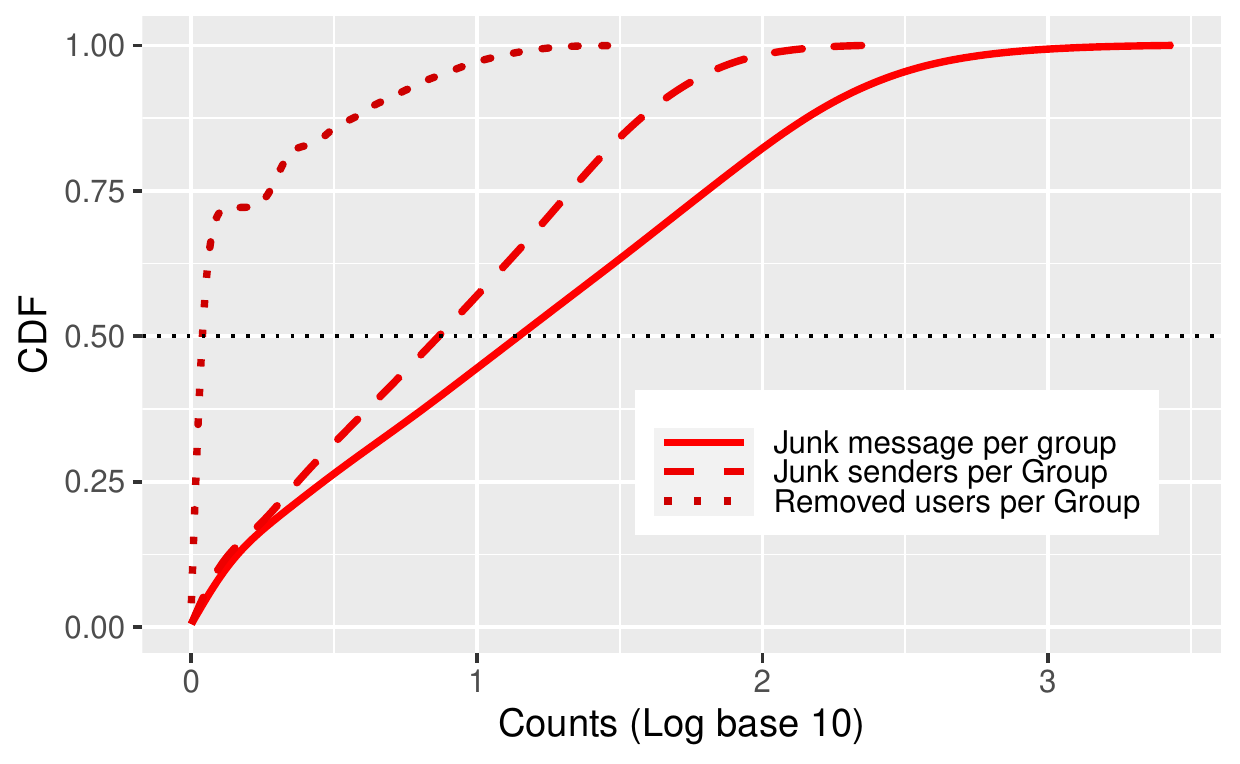}
    \caption{Count of junk messages, junk senders and removed users per group.}
    \label{fig:junkpergroup}
    \vspace{-\baselineskip}
\end{figure}

We start by inspecting the distribution of junk senders across the groups. 
Figure~\ref{fig:junkpergroup} presents the distribution of \one~junk messages per group; \two~junk senders per group; as well as \three~the number of users removed per group. First, we see that 75\% of groups have at least one junk sender, with an average of 13 per group (sending 58 junk messages per group). 
In the most extreme case, we observe 2722 junk messages in a single group. Overall, the top 25\% of groups have at least at least 18 junk senders.
This confirms that a large fraction of groups suffer with the accumulation of junk messages.
Figure~\ref{fig:junkpergroup} also shows that many groups actively filter junk senders: 17\% of groups have at least one example of a junk sender being removed. 
In fact, 2\% of groups (around 95) have removed at least 8 users.

\subsection{Junk sender locations}
\label{sec:spammer_locations}

\begin{figure}[t]
\centering
  \includegraphics[width=0.9\columnwidth]{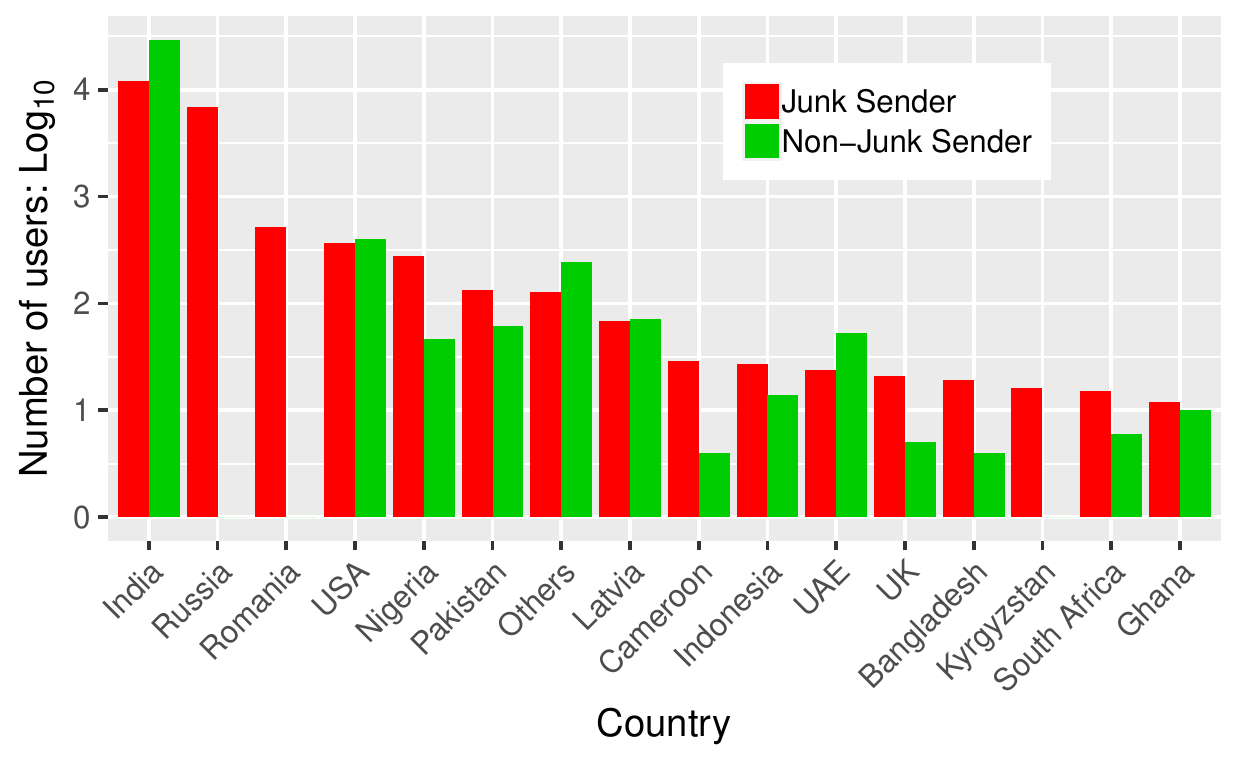}
 \caption{Country codes used by junk \vs non-junk senders}
\label{fig:international}
\vspace{-0.5cm}
\end{figure}

We next use the phone number country codes to geolocate all users. Figure~\ref{fig:international} presents the results for both junk and non-junk senders.
Unsurprisingly, the majority have Indian country codes, although we also see a range of other countries. Interestingly, these third party countries tend to be primarily junk senders.
Most striking is Russia, which has 6823 junk senders yet \emph{no} non-junk senders.
These users exclusively post junk content (9029 messages spread across 334 text clusters).
We also note that all Russian users' junk messages are in English.
Similar patterns are seen in phones from other countries, albeit on a smaller scale (\eg Romania, Kyrgyzstan).

\subsection{Longevity of junk messages}
\label{sec:coordination}

We now explore the temporal patterns of junk messages.
Recall that a cluster refers to group of similar messages (Section~\ref{sec:preprocessing}).
We term any day when at least 10 messages within a cluster are sent as an \textit{active day} for that particular cluster.

\begin{figure}[t]
\centering
\includegraphics[width=0.8\columnwidth]{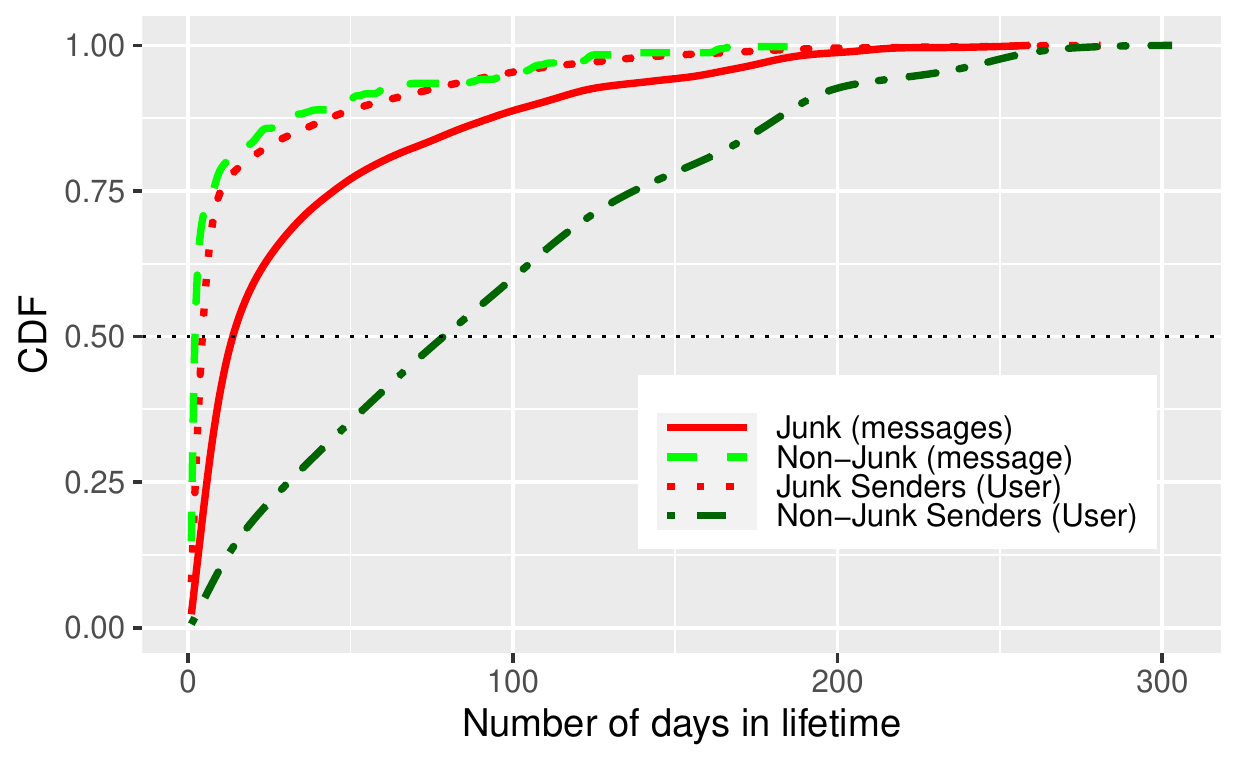}
  \caption{Cumulative Distribution Function (CDF) of lifetimes of junk and junk message clusters and users.}
  \label{fig:firstLastCDF}
  \vspace{-\baselineskip}
\end{figure}

We start by inspecting the lifetime of junk \vs non-junk messages.
We measure this as the difference between their first and last occurrence, shown in Figure \ref{fig:firstLastCDF}.
We see that junk messages have consistently longer lifetimes than non-junk. The median campaign duration for junk is 29 days compared to 12 days for non-junk.
Interestingly, we see the opposite trend when computing the lifetime of senders. 
Here, we compute the lifetime of a user as the difference between their first and last post (calculated across the user's activity in all groups in our dataset). As shown in Figure \ref{fig:firstLastCDF}, non-junk senders have substantially longer lifetimes than their junk senders counterparts, partly due to their removal from groups by admins. \looseness=-1

\begin{figure}[t!]
\centering
  \includegraphics[width=\columnwidth]{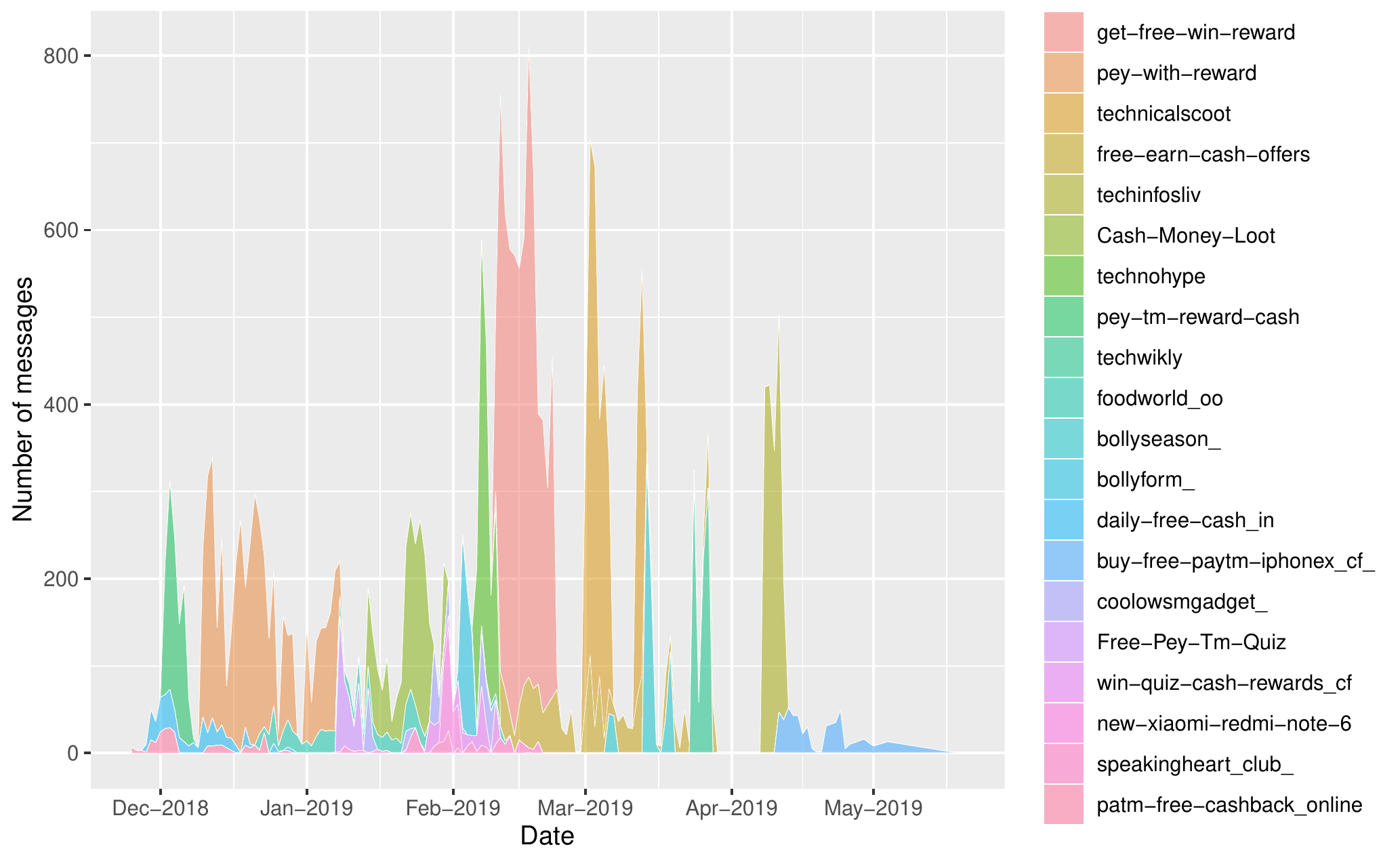}
  \includegraphics[width=0.49\columnwidth]{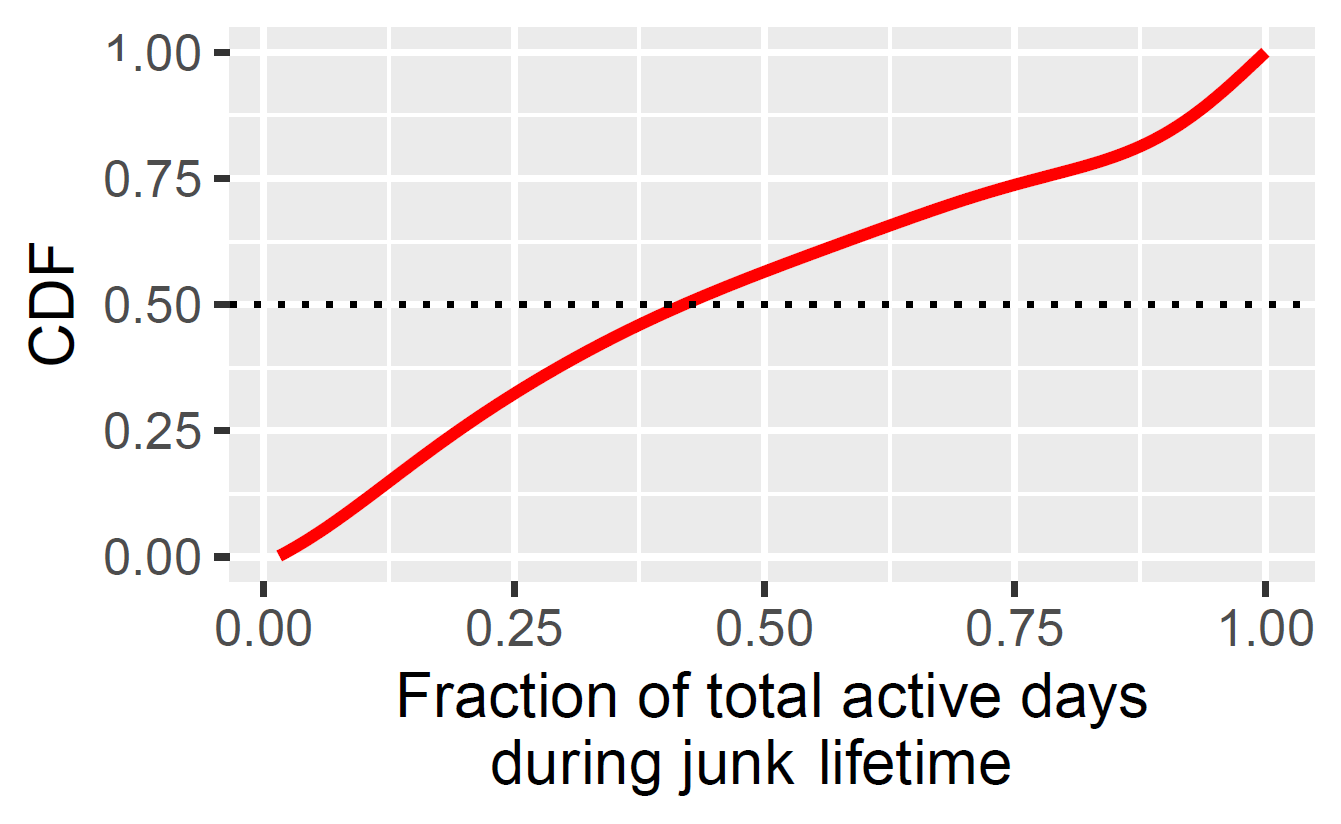}
 \includegraphics[width=0.49\columnwidth]{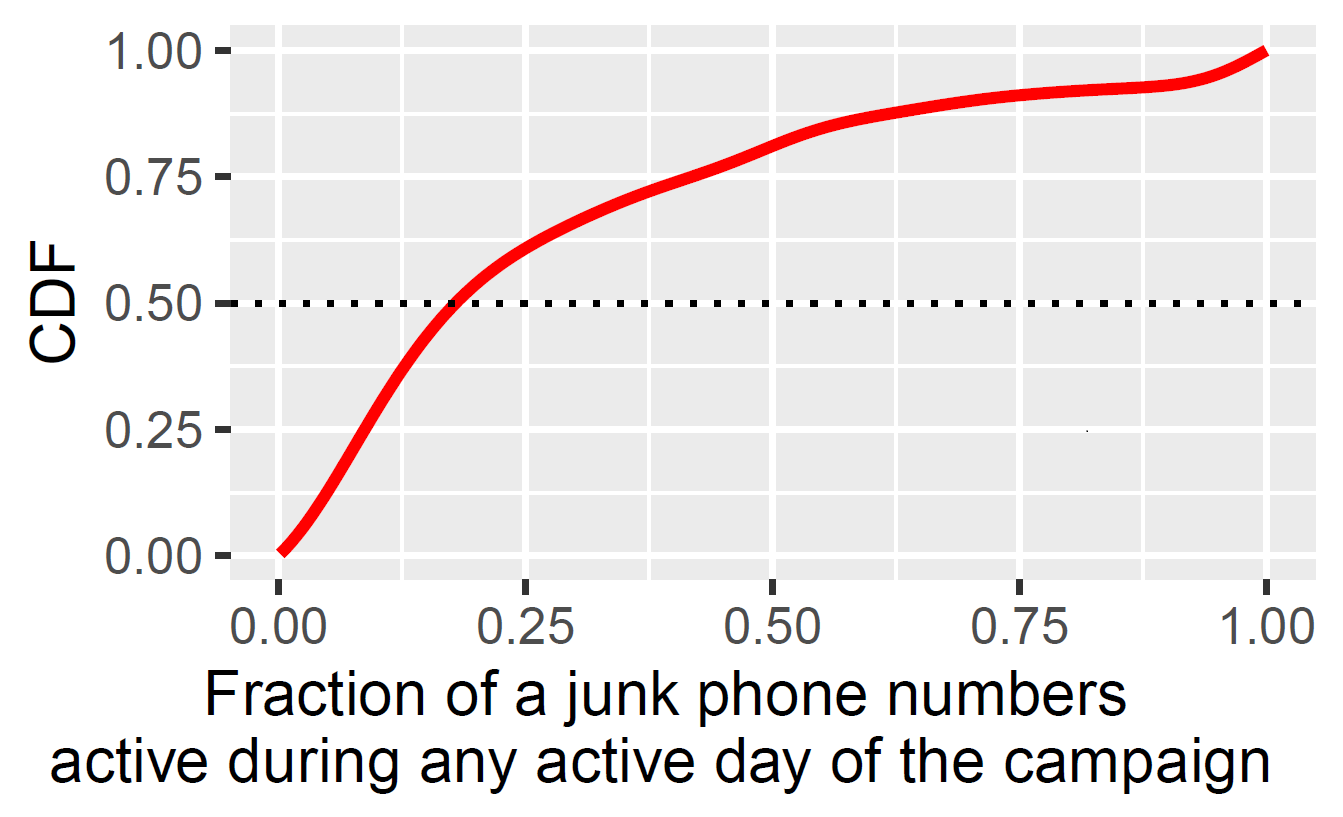}
 \caption{\textbf{Top}: Number of times a day that a junk message is posted, for 15 exemplar clusters. \textbf{Bottom Left}: Fraction of days that are `active' ($>$ 10 junk messages sent) during the lifetime of multi-day junk clusters. \textbf{Bottom Right}: Fraction of junk phone numbers used during active days.}
\label{fig:repeatedUsers}
\vspace{-\baselineskip}
\end{figure}

To gain a better understanding of these temporal trends we show 15 highly occurring example junk message from a single cluster in Figure \ref{fig:repeatedUsers} (top). These messages are repeatedly sent across multiple days and are highly focused, with aggressive peaks on a small number of days surrounded with different URLs.
To generalise this across all junk campaigns, Figure \ref{fig:repeatedUsers} (bottom left) plots the number of active days seen during the lifetime each cluster. This shows that most are sent across relatively few active days: On average less than half the days during the lifetime of a junk cluster are active with 10 or more messages.
This suggests junk senders take a staggered approach, rather than issuing junk messages every day. Despite this, there are a notable set of highly active messaging campaigns where messages are sent on most days: 25\% of clusters involve sending messages on at least 80\% of the days during their lifetime. 

The above leads us to conjecture that multiple phone numbers may be involved in sending messages for each cluster. 
To explore this, Figure \ref{fig:repeatedUsers} (bottom right) shows the fraction of the phone numbers involved in sending messages from each cluster that are active each day. 
We observe that, indeed, messages contained within a cluster are sent from multiple phone numbers.
On average, under 20\% of the phone numbers are involved on any active day.

\subsection{Joining and leaving groups}
\label{sec:joinleavedynamics}

\begin{figure}[t]
\centering
  \includegraphics[width=0.9\columnwidth]{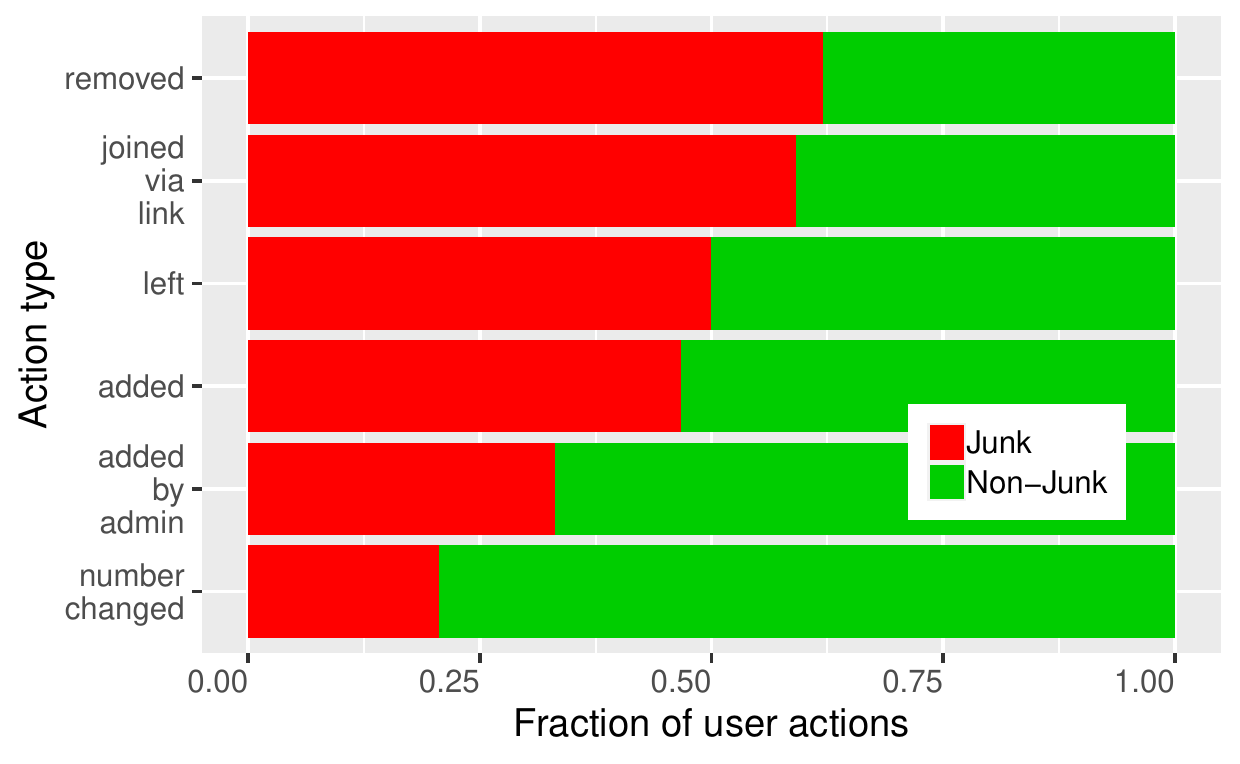}
  \caption{Relative proportions of join and leave actions.}
  \label{fig:spamJoinLeave}
  \vspace{-0.5cm}
\end{figure}

Finally, we examine how users join and leave different groups. 
Figure \ref{fig:spamJoinLeave} presents the fraction of junk senders \vs non-junk senders who join using each technique. 
A clear difference exists: it is more common for junk senders to join \wa groups via a link. Junk senders also comprise a disproportionate fraction of users who are `removed' from a group by the admin. 

\begin{table}
    \caption{Relation between leaving and joining methods for junk senders (equivalent numbers for non-junk senders in parenthesis).}
    \vspace{-12pt}
    \label{tab:joinLeaveDynamics}
    \centering
    \begin{tabular}{p{2.5cm}p{1.5cm}p{1.5cm}p{1.5cm}}
         \hline
        \textbf{Actions} & \textbf{Joined via link}& \textbf{Added}&
         \textbf{Added by admin} \\
         \hline
         \hline
         Left& 75\%~(62\%)& 17\%~(20\%)& 8\%~(18\%)\\
         \hline
         Number changed& 74\%~(37\%)& 12\%~(28\%)& 14\%~(35\%)\\
         \hline
         Removed& 80\%~(48\%)& 5\%~(18\%)& 15\%~(34\%)\\
         \hline
    \end{tabular}
\vspace{-0.5cm}
\end{table}

Table \ref{tab:joinLeaveDynamics} examines how users leave groups and how these actions are related to the method they use to join. Users who leave the group by any method are overwhelmingly likely to have joined via an invite link, although this is noticeably higher for junk senders than non-junk senders. Note that users who have been added by a group admin are the least likely to have left the group.

We can further inspect what users \emph{do} after joining. After joining via a link, nearly 40\% of junk senders (30\% of non-junk senders) post URLs. 18\% (5\%) post messages with a phone number, and 21\% (21\%) simply leave.
Again for junk senders, after posting a junk message with a URL, 86\% of the time the next action is to post another junk message with a URL.
We also check the actions of junk senders immediately before they are removed from a group. We find that 54\% of the time they posted a junk message with a URL, and 19\% of times they post a junk messages with a phone number in it. In total, 73\% of user removals by admins are immediately after the user posts a junk message.

\section{Junk Mitigation Strategies}
\label{sec:modeling}


In this section, we examine how we can automatically identify junk messages. The ultimate goal is to build a tool to \one~Help group admins identify junk senders who might need to be removed; and \two~Automatically filter such message proactively for users.

\begin{table*}[t]
    \caption{Junk detection performance using machine learning based classifiers.}
    \vspace{-12pt}
    \label{tab:accuracy}
    \centering
    \begin{tabular}{p{3.5cm}|p{3cm}|p{1.8cm}|p{1.7cm}|p{1.7cm}|p{1.5cm}}
       \hline
        \textbf{Features type}&\textbf{Model} &\textbf{Accuracy} &\textbf{F-1}& \textbf{Recall} & \textbf{Precision}
          \\
         \hline
         Content (Word Embedding)
         &Logistic Regression& 87.9\%&0.88&0.87&0.89\\
         \hline
         Content (Word Embedding)
         &SVM& 87.5\%&0.88&0.87&0.88\\
         \hline
         Content (Word Embedding)
         &Random Forest& \textbf{88.1}\%&0.88&0.88&0.89\\
         \hline
         \hline
         Metadata (All action+ISD)
         & Logistic Regression&67.5\%& 0.8& 0.67& 0.97\\
         \hline
         Metadata (All action+ISD)
         &SVM&74.4\%& 0.82 & 0.74 & 0.93 \\
         \hline
         Metadata (All action+ISD)
         &Random Forest &\textbf{87.5}\%& \textbf{0.90}& \textbf{0.87} & \textbf{0.94}\\
         \hline
         \hline
         Content+Action per group
         &Random Forest (Per group model)
         & 86\% (mean); sd=0.18
         & 0.63 (mean); sd=0.42
         &0.63 (mean); sd=0.44 &0.64 (mean); sd=0.43\\
    \hline
    \end{tabular}
\end{table*}

\subsection{Content-based junk detection}

We start by training a single global model using the entirety of our dataset.
To represent the text of messages, we rely on MuRIL (Multilingual Representations for Indian Languages) word embeddings~\cite{khanuja2021muril}. 
For each message in the 7k (3.5k junk) unique messages dataset that we label (Table~\ref{tab:annotation}), we obtain MuRIL embedding vectors of size 768, as described in~\cite{khanuja2021muril}. 
These 768 values constitute our input features.
We experiment with three binary classification models, for which the target is to classify messages into junk \vs non-junk: Logistic Regression (LR), Support Vector Machine (SVM) and Random Forest (RF).
We use a 80:20 train:test split using Sklearn library in Python 3. 
We tune the hyper-parameters in each model using Sklearn's GridSearchCV module.  Finally, we tune LR (C: 100 ; penalty:L1, solver=liblinear), SVM (kernal:rbf; C:100) and RF (number of estimators:400, min split:2; criterion: entropy), each with five-fold cross validation. 

The results are presented in Table~\ref{tab:accuracy} after testing with various combinations of penalty, estimators and optimisation criterion. We find that RF and SVM both perform well on embeddings (nearly 88\% accurate and 0.88 F-1 Score). 
This confirms that there \emph{are} sufficient determinant features to assist users with automated classification.

\subsection{Metadata-based junk detection by platform}
\label{sec:metadataclassifier}

A limitation of the above strategy is that training and classification requires access to raw text content. This, however, is challenging for the platform to implement due to \wa's end-to-end encryption.
Thus, models cannot easily be trained centrally, and text-based inference \emph{must} take place on the recipient's end device.
However, on-device ML may be challenging for low power mobile phones common in countries like India. \looseness=-1

With this in mind, we next experiment with an alternative approach that can be computed centrally by the \wa platform, without access to text content. Our key insight is that \textit{although content is encrypted, the actions of the users, such as joining or leaving groups are still visible}. That said, some of the most important features are use of phone numbers and URLs in text, which are contained in over 90\% of junk messages (\S\ref{sec:urls-phones}).  Thus, we propose a simple (optional) modification wherein each sender's \wa client encodes a 2-bit signal on whether the message sent contains a phone number, a URL, both or neither. The truthfulness of this signal can be verified by the recipients after decrypting, and the signal (though not the actual phone number or URL) can be made visible to the platform without compromising privacy.

Using this, we centrally build \emph{user profiles} upon which classification can be performed.
This contains each user's actions per-group as a vector (or features) of counts for the different types of actions (\eg number of times they joined/left a group). Table \ref{tab:featimp} (column 1) provides the full list of features.
Using these features, we again try a number of different methods including Logistic Regression (LR), Support Vector Machine (SVM) and Random Forest (RF). As a dataset, we consider all users with at least 2 actions. This leaves us with 47K user profiles with 15K removals and 32K non-removals across 3.6K groups. We again perform a random 80:20 five-fold cross-validation split for training and testing.
We tune hyper-parameters using the same approach as in the previous subsection, resulting in LR (C=0.001, solver='sag'; penalty:L2), SVM (kernal:rbf; C:1000) and RF (number of estimators: 500, min split: 2; criterion: entropy).

Our metadata-based classifier achieves similar performance to that of content-based modelling, with an F1-score (for best RF model) of 0.9 and accuracy of 87.5\% (Table \ref{tab:accuracy}). 
We also compare against an alternative model where our classifier does not have access to the 2 bit signal proposed above (this prevents the classifier from checking the presence of URLs or phone numbers). This decreases the F1-score to 0.67 and the accuracy to 0.82, indicating that even without this adaptation, our approach still can flag the majority of junk senders.
When inspecting feature importance, we observe that the most important feature is the number of messages posted (0.52 for RF), followed by the use of a non-domestic phone number (0.15 for RF). In cases where the 2-bit signal is not available the use of international phone number becomes the most important feature (0.42 for RF).\looseness=-1

\subsection{Content \& metadata detection by device}

Finally, we revisit the idea of performing local classification on the end device, such that we can use \emph{both} text content and metadata features simultaneously without undermining end-to-end encryption. 
This is possible because each user can gain vantage on the text content contained within their own groups. 
Thus, for each group, we construct a local on-device profile for each participating members.
This user profile contains all the features shown in Table~\ref{tab:featimp}, as well as the fraction of messages sent that are tagged as junk by content classifier. 
Importantly, these features can be locally constructed by any user in the group and do not require  central computation. 

\begin{figure}[t]
\centering
\includegraphics[width=0.8\columnwidth]{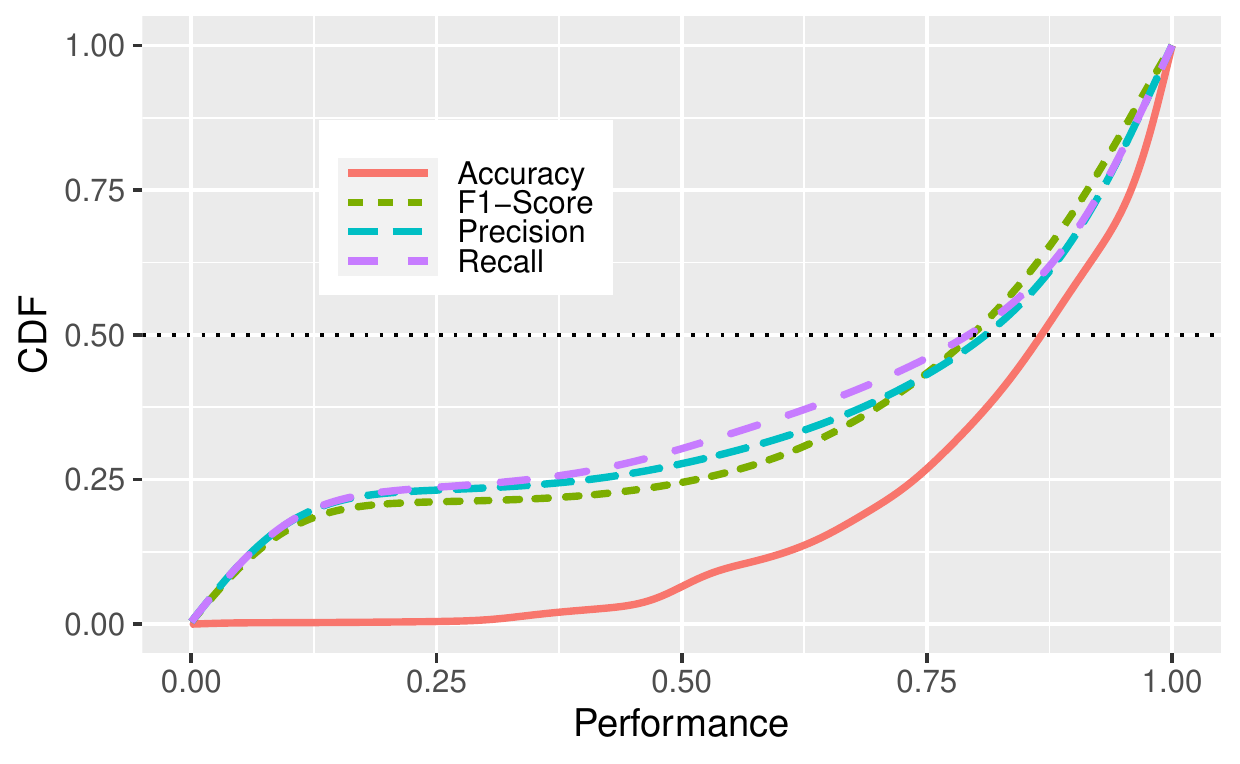}
  \caption{CDF of metrics for per-group Random Forest models.\looseness=-1 }
  \label{fig:localModelMetrix}
  
\end{figure}

We train a local Random Forest model on a \emph{per-group} basis (using the 3/4 of groups in our data that contain at least one junk sender). 
Here, we assume that supervised labels for local training can be obtained by annotation (\eg by the group admin). Our results show that only a small number of training samples would need to be collectively tagged by the admin and/or end users (a few hundred).

In our experiment, we use 80\% of users in each group for training, and 20\% for testing, tuned similarly as shown in models in previous sub-section. 
Table~\ref{tab:accuracy} summarises the results. We obtain 86\% mean accuracy across groups, which is close to the global model (0.9). 
Figure~\ref{fig:localModelMetrix} shows the distribution of performance scores across each \wa group. A minority of groups obtain poor performance, largely due to the small number of users.
For groups that gain under 60\% accuracy, we see just 300 messages and 35 users on average. Such groups contain insufficient data to effectively train the local models.
That said, 77\% of groups obtain accuracy exceeding 75\%. These tend to have larger populations, allowing end devices to better learn.
We therefore confirm that our classifier \emph{can} assist admins in identifying junk senders to remove, and assist users in proactively filtering such messages simultaneously.

\vspace{-12pt}
\section{Related Work}

Several studies have explored messaging patterns within other community mediums, \eg Reddit~\cite{singer2014evolution}, 4chan~\cite{bernstein20114chan,hine2017kek} and IRC~\cite{rintel2001first}.
Although there also have been studies of junk sending on social media~\cite{stringhini2010detecting,gao2010detecting,yardi2010detecting}, email~\cite{cormack2008email} and SMS~\cite{pervaiz2019assessment}, junk has not yet been studied on \wa except for anecdotal observations~\cite{threadReader, instagramSpam, indiatodayBan, whatsappban,agarwal2020characterising}. 
These studies have largely focussed on qualitative methodologies, \eg interviews, surveys, focus groups. For example, \cite{battestini2010large} collected quantitative data via the installation of a logging tool on user devices. By recruiting 70 participants, they analysed 58K sent messages. Although powerful, this approach is largely non-scalable.  Other messaging apps, such as WeChat~\cite{huang2015fine}, have been explored at scale although the focus has not been on the content and interactions. Instead, coarser analyses have been performed, \eg size of messages. Studies that have explored more social features have, again, limited themselves to small-scale surveys~\cite{lien2014examining}. 

There have been prior studies on the misuse of messaging services more generally, \eg SMS fraud in Pakistan~\cite{pervaiz2019assessment}.
There are also a small set of related works looking at the dissemination of malicious content via \wa groups. These, however, are largely focused on misinformation~\cite{javed2020first,javed2022deep}. Resende et al.~\cite{resende2019mis} investigated the dissemination of misinformation during the Brazilian elections, and the impact of introducing limits on message forwarding~\cite{de2019can}. In another similar study of Brazilian elections, Victor \etal   ~\cite{bursztyn2019thousands} found partisan activities in political groups.  Reis~\cite{reis2020can} proposed an architecture to flag misinformation in \wa without breaking end-to-end encryption. Unlike our approach, this relies on a manually annotated set of image hashes.  

In studying the presence of misuse, there have been a works attempting to detect  unsolicited spam campaigns and click-baits, primarily via emails or social media. For instance, Xiao et al.~\cite{xiao2015detecting} used supervised learning methods to detect groups of fake spam accounts on social media. 
This include tools for detecting unwanted videos in social media~\cite{benevenuto2009detecting,10.1145/3243734.3243770,10.1145/3319535.3345658,shin2021twiti}.
Boykin and Roychowdhury~\cite{boykin2005leveraging} investigated the use of social graphs to filter spam. 
There are various content-based approaches to detecting junk too~\cite{balli2018development,almeida2011contributions,ma2016intelligent}. These tend to rely on building document models and training machine classifiers (\eg SVMs) to detect spam messages. 
In the case of \wa, this can be problematic due to its use of end-to-end encryption. We consider such studies orthogonal to our work.


\section{Conclusions}

This paper has presented the first study of junk messages in public \wa groups. We have gathered 2.6 million messages from 5,051 public politics-related groups in India, and analysed the content, URLs and temporal patterns of spam.
We find that junk is commonplace on \wa, and senders tend to post across a large number of groups. They also exhibit interesting patterns of leaving and rejoining groups multiple times, to avoid being removed by admins.
We further find evidence of campaigns --- spreading the same spam message over a small number of `active' days. These strategies may help improve the visibility of junk, by providing a longer `shelf life' in the recent messages.\looseness=-1

Our results have clear implications for junk detection. For example, a key indicator of junk is the presence of particular URLs and phone numbers. We have shown that this can be used as part of automated detection, and have demonstrated that models can be trained to detect \wa junk. 
This can assist admins in quickly flagging and removing such users.
However, as \wa uses end-to-end encryption, such information cannot be accessed by the platform. We have therefore proposed techniques that can be used by the end device or the platform (centrally), whilst still respecting end-to-end encryption guarantees. 
To aid reproducibility, our annotated dataset and code is publicly
available for researchers.\footnote{More information is available at: \url{http://tiny.cc/netsys-whats-app}}


\subsection*{Acknowledgements}

We thank the annotators from India. This work is partially supported by EU H2020 grant agreements No 830927 (Concordia) and No 101016509 (Charity). Gareth Tyson was supported by grant EP/S033564/1. Kiran Garimella did the work at MIT, supported by a Michael Hammer Postdoctoral Fellowship.

\section*{Appendix}
\appendix



\subsection*{Example Junk Messages}

\begin{figure}[h]
    \centering
    \includegraphics[width=0.9\columnwidth]{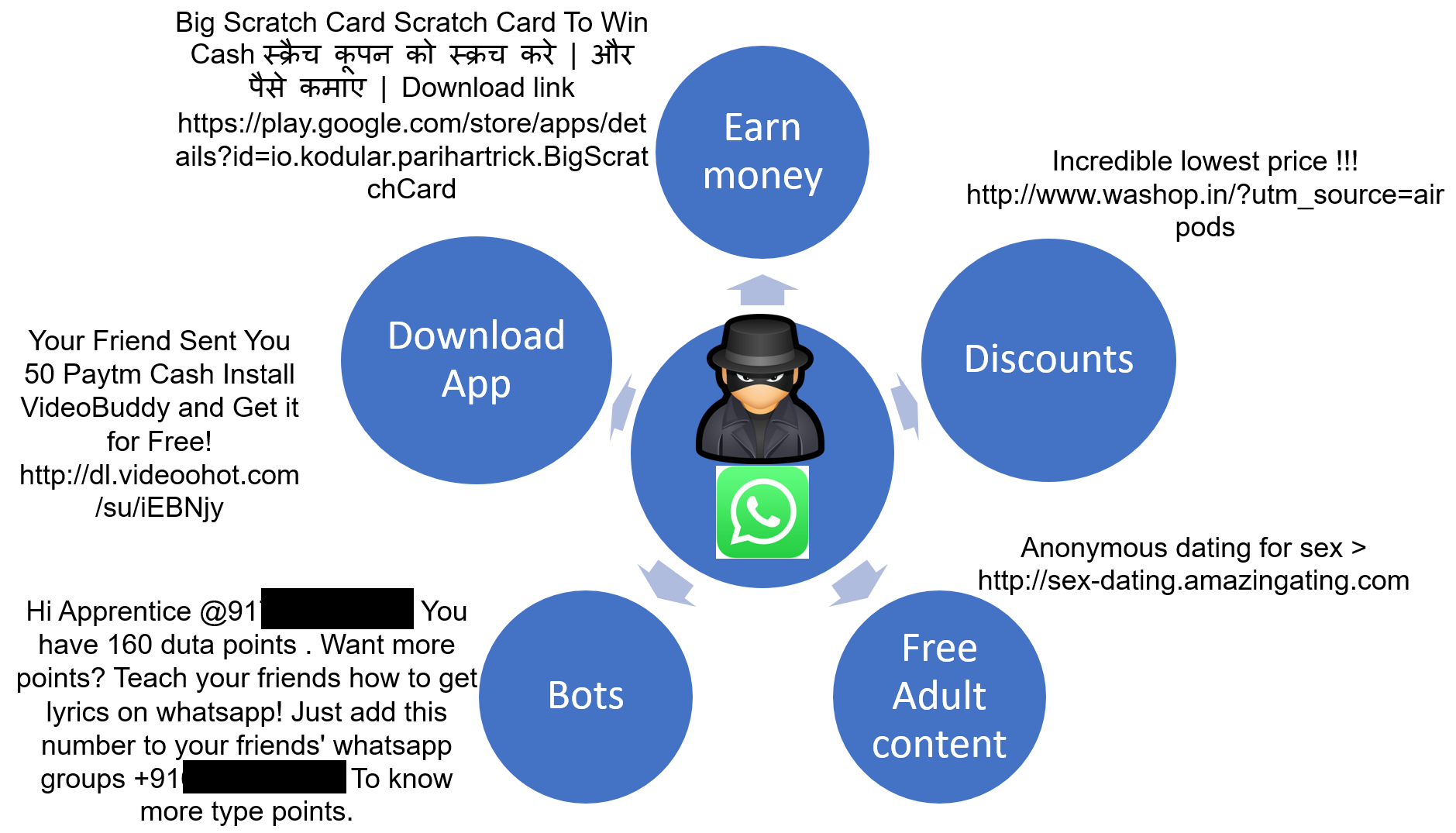}
    \caption{Some top examples of junk messages.}
    \label{fig:typesSpam}
\end{figure}

Figure~\ref{fig:typesSpam} shows several examples of junk messages. Table~\ref{tab:featimp} summarises the most significant features in our Random Forest Classifier (see \S\ref{sec:metadataclassifier}).

\vspace{15pt}

\begin{table}[h]
    \caption{Feature importance of Random Forest Classifier to separate non-jettisons and jettisons.}
    \label{tab:featimp}
    \centering
    \resizebox{\columnwidth}{!}{
    \begin{tabular}{c|c|c}
        \hline
    \textbf{Feature} & \textbf{With 2-bit signal} & \textbf{No 2-bit signal}\\
    \hline
    \hline
    Posted simple message    & 0.52& 0.37\\
    Non-domestic number   & 0.15 &0.42\\
    Posted URL    & 0.12 &N/A\\
    Joined via link    &  0.08&0.075\\
    Posted phone number     &0.05&  N/A\\
    Left group   & 0.04 &0.05\\
    Added by member    &  0.023&0.03\\
    Added by admin    & 0.021& 0.025\\
    Removed from group    & 0.01&0.015 \\
    Number changed    &0.003  &0.003\\
    \hline
    \end{tabular}    }
    \vspace{-\baselineskip}
\end{table}

\clearpage

\bibliographystyle{ACM-Reference-Format}
\bibliography{sample-base}


\begin{thebibliography}{49}


\ifx \showCODEN    \undefined \def \showCODEN     #1{\unskip}     \fi
\ifx \showDOI      \undefined \def \showDOI       #1{#1}\fi
\ifx \showISBNx    \undefined \def \showISBNx     #1{\unskip}     \fi
\ifx \showISBNxiii \undefined \def \showISBNxiii  #1{\unskip}     \fi
\ifx \showISSN     \undefined \def \showISSN      #1{\unskip}     \fi
\ifx \showLCCN     \undefined \def \showLCCN      #1{\unskip}     \fi
\ifx \shownote     \undefined \def \shownote      #1{#1}          \fi
\ifx \showarticletitle \undefined \def \showarticletitle #1{#1}   \fi
\ifx \showURL      \undefined \def \showURL       {\relax}        \fi
\providecommand\bibfield[2]{#2}
\providecommand\bibinfo[2]{#2}
\providecommand\natexlab[1]{#1}
\providecommand\showeprint[2][]{arXiv:#2}

\bibitem[\protect\citeauthoryear{Agarwal, Garimella, Joglekar, Sastry, and
  Tyson}{Agarwal et~al\mbox{.}}{2020}]%
        {agarwal2020characterising}
\bibfield{author}{\bibinfo{person}{Pushkal Agarwal}, \bibinfo{person}{Kiran
  Garimella}, \bibinfo{person}{Sagar Joglekar}, \bibinfo{person}{Nishanth
  Sastry}, {and} \bibinfo{person}{Gareth Tyson}.}
  \bibinfo{year}{2020}\natexlab{}.
\newblock \showarticletitle{Characterising user content on a multi-lingual
  social network}. In \bibinfo{booktitle}{{\em ICWSM}}.
\newblock


\bibitem[\protect\citeauthoryear{Almeida, Hidalgo, and Yamakami}{Almeida
  et~al\mbox{.}}{2011}]%
        {almeida2011contributions}
\bibfield{author}{\bibinfo{person}{Tiago~A Almeida}, \bibinfo{person}{Jos{\'e}
  Mar{\'\i}a~G Hidalgo}, {and} \bibinfo{person}{Akebo Yamakami}.}
  \bibinfo{year}{2011}\natexlab{}.
\newblock \showarticletitle{Contributions to the study of SMS spam filtering:
  new collection and results}. In \bibinfo{booktitle}{{\em Symposium on
  Document engineering}}.
\newblock


\bibitem[\protect\citeauthoryear{Ananth and Sharma}{Ananth and Sharma}{2019}]%
        {instagramSpam}
\bibfield{author}{\bibinfo{person}{Venkat Ananth} {and}
  \bibinfo{person}{Samidha Sharma}.} \bibinfo{year}{2019}\natexlab{}.
\newblock \bibinfo{title}{On Instagram, in India, it’s sex for sale}.
\newblock \bibinfo{howpublished}{The Economic Times}.   (\bibinfo{year}{2019}).
\newblock
\showURL{%
\url{https://economictimes.indiatimes.com/tech/internet/on-instagram-in-india-its-sex-for-sale/articleshow/67968645.cms?from=mdr}}


\bibitem[\protect\citeauthoryear{Ball{\i} and Karasoy}{Ball{\i} and
  Karasoy}{2018}]%
        {balli2018development}
\bibfield{author}{\bibinfo{person}{Serkan Ball{\i}} {and} \bibinfo{person}{Onur
  Karasoy}.} \bibinfo{year}{2018}\natexlab{}.
\newblock \showarticletitle{Development of content-based SMS classification
  application by using Word2Vec-based feature extraction}.
\newblock \bibinfo{journal}{{\em IET Software\/}} (\bibinfo{year}{2018}).
\newblock


\bibitem[\protect\citeauthoryear{Banaji et~al\mbox{.}}{Banaji
  et~al\mbox{.}}{2019}]%
        {banaji2019whatsapp}
\bibfield{author}{\bibinfo{person}{Shakuntala Banaji} {et~al\mbox{.}}}
  \bibinfo{year}{2019}\natexlab{}.
\newblock \showarticletitle{WhatsApp vigilantes: an exploration of citizen
  reception and circulation of WhatsApp misinformation linked to mob violence
  in India}.
\newblock \bibinfo{journal}{{\em Dept.\ of Media and Communications, LSE\/}}
  (\bibinfo{year}{2019}).
\newblock


\bibitem[\protect\citeauthoryear{Battestini, Setlur, and Sohn}{Battestini
  et~al\mbox{.}}{2010}]%
        {battestini2010large}
\bibfield{author}{\bibinfo{person}{Agathe Battestini}, \bibinfo{person}{Vidya
  Setlur}, {and} \bibinfo{person}{Timothy Sohn}.}
  \bibinfo{year}{2010}\natexlab{}.
\newblock \showarticletitle{A large scale study of text-messaging use}. In
  \bibinfo{booktitle}{{\em Proceedings of the 12th international conference on
  Human computer interaction with mobile devices and services}}. ACM.
\newblock


\bibitem[\protect\citeauthoryear{Benevenuto et~al\mbox{.}}{Benevenuto
  et~al\mbox{.}}{2009}]%
        {benevenuto2009detecting}
\bibfield{author}{\bibinfo{person}{Fabr{\'\i}cio Benevenuto} {et~al\mbox{.}}}
  \bibinfo{year}{2009}\natexlab{}.
\newblock \showarticletitle{Detecting spammers and content promoters in online
  video social networks}. In \bibinfo{booktitle}{{\em ACM SIGIR}}.
\newblock


\bibitem[\protect\citeauthoryear{Bernstein, Monroy-Hern{\'a}ndez, Harry,
  Andr{\'e}, Panovich, and Vargas}{Bernstein et~al\mbox{.}}{2011}]%
        {bernstein20114chan}
\bibfield{author}{\bibinfo{person}{Michael~S Bernstein},
  \bibinfo{person}{Andr{\'e}s Monroy-Hern{\'a}ndez}, \bibinfo{person}{Drew
  Harry}, \bibinfo{person}{Paul Andr{\'e}}, \bibinfo{person}{Katrina Panovich},
  {and} \bibinfo{person}{Gregory~G Vargas}.} \bibinfo{year}{2011}\natexlab{}.
\newblock \showarticletitle{4chan and /b/: An Analysis of Anonymity and
  Ephemerality in a Large Online Community.}. In \bibinfo{booktitle}{{\em
  ICWSM}}.
\newblock


\bibitem[\protect\citeauthoryear{Boykin and Roychowdhury}{Boykin and
  Roychowdhury}{2005}]%
        {boykin2005leveraging}
\bibfield{author}{\bibinfo{person}{P~Oscar Boykin} {and}
  \bibinfo{person}{Vwani~P Roychowdhury}.} \bibinfo{year}{2005}\natexlab{}.
\newblock \showarticletitle{Leveraging social networks to fight spam}.
\newblock \bibinfo{journal}{{\em IEEE\/}} (\bibinfo{year}{2005}).
\newblock


\bibitem[\protect\citeauthoryear{Bureau}{Bureau}{2021}]%
        {whatsappban}
\bibfield{author}{\bibinfo{person}{Gadgets~Now Bureau}.}
  \bibinfo{year}{2021}\natexlab{}.
\newblock \bibinfo{title}{WhatsApp bans Indian accounts}.
\newblock   (\bibinfo{year}{2021}).
\newblock
\newblock
\shownote{\url{bit.ly/3nyr0br}.}


\bibitem[\protect\citeauthoryear{Bursztyn and Birnbaum}{Bursztyn and
  Birnbaum}{2019}]%
        {bursztyn2019thousands}
\bibfield{author}{\bibinfo{person}{Victor~S Bursztyn} {and}
  \bibinfo{person}{Larry Birnbaum}.} \bibinfo{year}{2019}\natexlab{}.
\newblock \showarticletitle{Thousands of small, constant rallies: A large-scale
  analysis of partisan WhatsApp groups}. In \bibinfo{booktitle}{{\em ASONAM}}.
  IEEE.
\newblock


\bibitem[\protect\citeauthoryear{Cormack}{Cormack}{2008}]%
        {cormack2008email}
\bibfield{author}{\bibinfo{person}{Gordon~V Cormack}.}
  \bibinfo{year}{2008}\natexlab{}.
\newblock \bibinfo{booktitle}{{\em Email spam filtering: A systematic review}}.
\newblock \bibinfo{publisher}{Now Publishers}.
\newblock


\bibitem[\protect\citeauthoryear{de~Freitas~Melo et~al\mbox{.}}{de~Freitas~Melo
  et~al\mbox{.}}{2019}]%
        {de2019can}
\bibfield{author}{\bibinfo{person}{de Freitas~Melo} {et~al\mbox{.}}}
  \bibinfo{year}{2019}\natexlab{}.
\newblock \showarticletitle{Can WhatsApp Counter Misinformation by Limiting
  Message Forwarding?}. In \bibinfo{booktitle}{{\em Conf.\ Compl. Netw.\ and
  Appl.}}
\newblock


\bibitem[\protect\citeauthoryear{{DUTA}}{{DUTA}}{2018}]%
        {duta_bot}
\bibfield{author}{\bibinfo{person}{{DUTA}}.} \bibinfo{year}{2018}\natexlab{}.
\newblock \bibinfo{title}{Duta.in: BRINGING THE INTERNET TO THE NEXT BILLION}.
\newblock   (\bibinfo{year}{2018}).
\newblock
\newblock
\shownote{\url{duta.in}.}


\bibitem[\protect\citeauthoryear{Gao et~al\mbox{.}}{Gao et~al\mbox{.}}{2010}]%
        {gao2010detecting}
\bibfield{author}{\bibinfo{person}{Hongyu Gao} {et~al\mbox{.}}}
  \bibinfo{year}{2010}\natexlab{}.
\newblock \showarticletitle{Detecting and characterizing social spam
  campaigns}. In \bibinfo{booktitle}{{\em SIGCOMM IMC}}.
\newblock


\bibitem[\protect\citeauthoryear{Garimella and Tyson}{Garimella and
  Tyson}{2018}]%
        {garimella2018whatsapp}
\bibfield{author}{\bibinfo{person}{Kiran Garimella} {and}
  \bibinfo{person}{Gareth Tyson}.} \bibinfo{year}{2018}\natexlab{}.
\newblock \showarticletitle{Whatapp doc? A first look at whatsapp public group
  data}. In \bibinfo{booktitle}{{\em ICWSM}}.
\newblock


\bibitem[\protect\citeauthoryear{Gionis, Indyk, Motwani, et~al\mbox{.}}{Gionis
  et~al\mbox{.}}{1999}]%
        {gionis1999similarity}
\bibfield{author}{\bibinfo{person}{Aristides Gionis}, \bibinfo{person}{Piotr
  Indyk}, \bibinfo{person}{Rajeev Motwani}, {et~al\mbox{.}}}
  \bibinfo{year}{1999}\natexlab{}.
\newblock \showarticletitle{Similarity search in high dimensions via hashing}.
  In \bibinfo{booktitle}{{\em VLDB}}.
\newblock


\bibitem[\protect\citeauthoryear{Gupta, Singh, Ahuja, and Gupta}{Gupta
  et~al\mbox{.}}{2019}]%
        {gupta2019good}
\bibfield{author}{\bibinfo{person}{Aakriti Gupta}, \bibinfo{person}{Sunil~Kumar
  Singh}, \bibinfo{person}{Kabir Ahuja}, {and} \bibinfo{person}{Ankit Gupta}.}
  \bibinfo{year}{2019}\natexlab{}.
\newblock \showarticletitle{Good Morning Turning to Spam Morning}. In
  \bibinfo{booktitle}{{\em ICICCT}}.
\newblock


\bibitem[\protect\citeauthoryear{Hernandez, Rahman, Recabarren, and
  Carbunar}{Hernandez et~al\mbox{.}}{2018}]%
        {10.1145/3243734.3243770}
\bibfield{author}{\bibinfo{person}{Nestor Hernandez}, \bibinfo{person}{Mizanur
  Rahman}, \bibinfo{person}{Ruben Recabarren}, {and} \bibinfo{person}{Bogdan
  Carbunar}.} \bibinfo{year}{2018}\natexlab{}.
\newblock \showarticletitle{Fraud De-Anonymization for Fun and Profit} {\em
  (\bibinfo{series}{CCS})}. \bibinfo{publisher}{ACM}.
\newblock


\bibitem[\protect\citeauthoryear{Hine, Onaolapo, De~Cristofaro, Kourtellis,
  Leontiadis, Samaras, Stringhini, and Blackburn}{Hine et~al\mbox{.}}{2017}]%
        {hine2017kek}
\bibfield{author}{\bibinfo{person}{Gabriel~Emile Hine},
  \bibinfo{person}{Jeremiah Onaolapo}, \bibinfo{person}{Emiliano
  De~Cristofaro}, \bibinfo{person}{Nicolas Kourtellis}, \bibinfo{person}{Ilias
  Leontiadis}, \bibinfo{person}{Riginos Samaras}, \bibinfo{person}{Gianluca
  Stringhini}, {and} \bibinfo{person}{Jeremy Blackburn}.}
  \bibinfo{year}{2017}\natexlab{}.
\newblock \showarticletitle{Kek, Cucks, and God Emperor Trump: A Measurement
  Study of 4chan's Politically Incorrect Forum and Its Effects on the Web.}. In
  \bibinfo{booktitle}{{\em ICWSM}}.
\newblock


\bibitem[\protect\citeauthoryear{Huang, Lee, He, Qian, and He}{Huang
  et~al\mbox{.}}{2015}]%
        {huang2015fine}
\bibfield{author}{\bibinfo{person}{Qun Huang}, \bibinfo{person}{Patrick~PC
  Lee}, \bibinfo{person}{Caifeng He}, \bibinfo{person}{Jianfeng Qian}, {and}
  \bibinfo{person}{Cheng He}.} \bibinfo{year}{2015}\natexlab{}.
\newblock \showarticletitle{Fine-grained dissection of WeChat in cellular
  networks}. In \bibinfo{booktitle}{{\em Quality of Service (IWQoS), 2015 IEEE
  23rd International Symposium on}}. IEEE.
\newblock


\bibitem[\protect\citeauthoryear{Ikram, Masood, Tyson, Kaafar, Loizon, and
  Ensafi}{Ikram et~al\mbox{.}}{2019}]%
        {ikram2019chain}
\bibfield{author}{\bibinfo{person}{Muhammad Ikram}, \bibinfo{person}{Rahat
  Masood}, \bibinfo{person}{Gareth Tyson}, \bibinfo{person}{Mohamed~Ali
  Kaafar}, \bibinfo{person}{Noha Loizon}, {and} \bibinfo{person}{Roya Ensafi}.}
  \bibinfo{year}{2019}\natexlab{}.
\newblock \showarticletitle{The Chain of Implicit Trust: An Analysis of the Web
  Third-party Resources Loading}.
\newblock \bibinfo{journal}{{\em Web Conference\/}} (\bibinfo{year}{2019}).
\newblock


\bibitem[\protect\citeauthoryear{Javed, Shuja, Usama, Qadir, Iqbal, Tyson,
  Castro, and Garimella}{Javed et~al\mbox{.}}{2020}]%
        {javed2020first}
\bibfield{author}{\bibinfo{person}{R~Tallal Javed},
  \bibinfo{person}{Mirza~Elaaf Shuja}, \bibinfo{person}{Muhammad Usama},
  \bibinfo{person}{Junaid Qadir}, \bibinfo{person}{Waleed Iqbal},
  \bibinfo{person}{Gareth Tyson}, \bibinfo{person}{Ignacio Castro}, {and}
  \bibinfo{person}{Kiran Garimella}.} \bibinfo{year}{2020}\natexlab{}.
\newblock \showarticletitle{A First Look at COVID-19 Messages on WhatsApp in
  Pakistan}. In \bibinfo{booktitle}{{\em 2020 IEEE/ACM International Conference
  on Advances in Social Networks Analysis and Mining (ASONAM)}}. IEEE,
  \bibinfo{pages}{118--125}.
\newblock


\bibitem[\protect\citeauthoryear{Javed, Usama, Iqbal, Qadir, Tyson, Castro, and
  Garimella}{Javed et~al\mbox{.}}{2022}]%
        {javed2022deep}
\bibfield{author}{\bibinfo{person}{R~Tallal Javed}, \bibinfo{person}{Muhammad
  Usama}, \bibinfo{person}{Waleed Iqbal}, \bibinfo{person}{Junaid Qadir},
  \bibinfo{person}{Gareth Tyson}, \bibinfo{person}{Ignacio Castro}, {and}
  \bibinfo{person}{Kiran Garimella}.} \bibinfo{year}{2022}\natexlab{}.
\newblock \showarticletitle{A deep dive into COVID-19-related messages on
  WhatsApp in Pakistan}.
\newblock \bibinfo{journal}{{\em Social Network Analysis and Mining\/}}
  \bibinfo{volume}{12}, \bibinfo{number}{1} (\bibinfo{year}{2022}),
  \bibinfo{pages}{1--16}.
\newblock


\bibitem[\protect\citeauthoryear{Jones}{Jones}{2017}]%
        {jones2017whatsapp}
\bibfield{author}{\bibinfo{person}{Matt Jones}.}
  \bibinfo{year}{2017}\natexlab{}.
\newblock \bibinfo{title}{How WhatsApp Reduced Spam while Launching End-to-End
  Encryption}.
\newblock   (\bibinfo{year}{2017}).
\newblock
\showURL{%
\url{usenix.org/conference/enigma2017/conference-program/presentation/jones}}


\bibitem[\protect\citeauthoryear{Khanuja et~al\mbox{.}}{Khanuja
  et~al\mbox{.}}{2021}]%
        {khanuja2021muril}
\bibfield{author}{\bibinfo{person}{Simran Khanuja} {et~al\mbox{.}}}
  \bibinfo{year}{2021}\natexlab{}.
\newblock \bibinfo{title}{MuRIL: Multilingual Representations for Indian
  Languages}.
\newblock \bibinfo{howpublished}{Arxiv, tfhub.dev/google/MuRIL/1}.
  (\bibinfo{year}{2021}).
\newblock


\bibitem[\protect\citeauthoryear{Kim et~al\mbox{.}}{Kim et~al\mbox{.}}{2015}]%
        {kim2015detecting}
\bibfield{author}{\bibinfo{person}{Dae~Wook Kim} {et~al\mbox{.}}}
  \bibinfo{year}{2015}\natexlab{}.
\newblock \showarticletitle{Detecting fake anti-virus software distribution
  webpages}.
\newblock \bibinfo{journal}{{\em Computers \& Security\/}}
  (\bibinfo{year}{2015}).
\newblock


\bibitem[\protect\citeauthoryear{Lien and Cao}{Lien and Cao}{2014}]%
        {lien2014examining}
\bibfield{author}{\bibinfo{person}{Che~Hui Lien} {and} \bibinfo{person}{Yang
  Cao}.} \bibinfo{year}{2014}\natexlab{}.
\newblock \showarticletitle{Examining WeChat users’ motivations, trust,
  attitudes, and positive word-of-mouth: Evidence from China}.
\newblock \bibinfo{journal}{{\em Computers in Human Behavior\/}}
  (\bibinfo{year}{2014}).
\newblock


\bibitem[\protect\citeauthoryear{Lokniti}{Lokniti}{2018}]%
        {lokniti2018}
\bibfield{author}{\bibinfo{person}{CSDS Lokniti}.}
  \bibinfo{year}{2018}\natexlab{}.
\newblock \bibinfo{title}{How widespread is WhatsApp's usage in India?}
\newblock \bibinfo{howpublished}{Live Mint}.   (\bibinfo{year}{2018}).
\newblock
\showURL{%
\url{livemint.com/Technology/O6DLmIibCCV5luEG9XuJWL/How-widespread-is-WhatsApps-usage-in-India.html}}


\bibitem[\protect\citeauthoryear{Ma et~al\mbox{.}}{Ma et~al\mbox{.}}{2016}]%
        {ma2016intelligent}
\bibfield{author}{\bibinfo{person}{Jialin Ma} {et~al\mbox{.}}}
  \bibinfo{year}{2016}\natexlab{}.
\newblock \showarticletitle{Intelligent SMS spam filtering using topic model}.
  In \bibinfo{booktitle}{{\em Intl.\ Conf.\ on Intelligent Netw.\ and Collab.\
  Systems (INCoS)}}.
\newblock


\bibitem[\protect\citeauthoryear{Mehrotra, Pejovic, Vermeulen, Hendley, and
  Musolesi}{Mehrotra et~al\mbox{.}}{2016}]%
        {mehrotra2016my}
\bibfield{author}{\bibinfo{person}{Abhinav Mehrotra}, \bibinfo{person}{Veljko
  Pejovic}, \bibinfo{person}{Jo Vermeulen}, \bibinfo{person}{Robert Hendley},
  {and} \bibinfo{person}{Mirco Musolesi}.} \bibinfo{year}{2016}\natexlab{}.
\newblock \showarticletitle{My phone and me: understanding people's receptivity
  to mobile notifications}. In \bibinfo{booktitle}{{\em CHI}}.
\newblock


\bibitem[\protect\citeauthoryear{Mullen}{Mullen}{2015}]%
        {mullen2015textreuse}
\bibfield{author}{\bibinfo{person}{Lincoln Mullen}.}
  \bibinfo{year}{2015}\natexlab{}.
\newblock \showarticletitle{textreuse: Detect text reuse and document
  similarity}.
\newblock \bibinfo{journal}{{\em rOpenSci\/}} (\bibinfo{year}{2015}).
\newblock


\bibitem[\protect\citeauthoryear{Newman, Fletcher, Kalogeropoulos, and
  Nielsen}{Newman et~al\mbox{.}}{2019}]%
        {reuters2019report}
\bibfield{author}{\bibinfo{person}{Nic Newman}, \bibinfo{person}{Richard
  Fletcher}, \bibinfo{person}{Antonis Kalogeropoulos}, {and}
  \bibinfo{person}{Rasmus~Kleis Nielsen}.} \bibinfo{year}{2019}\natexlab{}.
\newblock \bibinfo{title}{{Reuters Institute Digital News Report 2019 }}.
\newblock   (\bibinfo{year}{2019}).
\newblock


\bibitem[\protect\citeauthoryear{Ooms and Sites}{Ooms and Sites}{2018}]%
        {ooms2018cld2}
\bibfield{author}{\bibinfo{person}{J Ooms} {and} \bibinfo{person}{D Sites}.}
  \bibinfo{year}{2018}\natexlab{}.
\newblock \showarticletitle{cld2: Google’s Compact Language Detector 2}.
\newblock  (\bibinfo{year}{2018}).
\newblock


\bibitem[\protect\citeauthoryear{Pathak}{Pathak}{2019}]%
        {indiatodayBan}
\bibfield{author}{\bibinfo{person}{Priya Pathak}.}
  \bibinfo{year}{2019}\natexlab{}.
\newblock \bibinfo{title}{WhatsApp is banning 2 million accounts every month}.
\newblock \bibinfo{howpublished}{India Today}.   (\bibinfo{year}{2019}).
\newblock
\showURL{%
\url{indiatoday.in/technology/features/story/whatsapp-is-banning-2-million-accounts}}


\bibitem[\protect\citeauthoryear{Pervaiz et~al\mbox{.}}{Pervaiz
  et~al\mbox{.}}{2019}]%
        {pervaiz2019assessment}
\bibfield{author}{\bibinfo{person}{Fahad Pervaiz} {et~al\mbox{.}}}
  \bibinfo{year}{2019}\natexlab{}.
\newblock \showarticletitle{An assessment of {SMS} fraud in Pakistan}. In
  \bibinfo{booktitle}{{\em ACM CCS}}.
\newblock


\bibitem[\protect\citeauthoryear{Pielot, Church, and De~Oliveira}{Pielot
  et~al\mbox{.}}{2014}]%
        {pielot2014situ}
\bibfield{author}{\bibinfo{person}{Martin Pielot}, \bibinfo{person}{Karen
  Church}, {and} \bibinfo{person}{Rodrigo De~Oliveira}.}
  \bibinfo{year}{2014}\natexlab{}.
\newblock \showarticletitle{An in-situ study of mobile phone notifications}. In
  \bibinfo{booktitle}{{\em International conference on Human-computer
  interaction}}.
\newblock


\bibitem[\protect\citeauthoryear{Rahman, Hernandez, Recabarren, Ahmed, and
  Carbunar}{Rahman et~al\mbox{.}}{2019}]%
        {10.1145/3319535.3345658}
\bibfield{author}{\bibinfo{person}{Mizanur Rahman}, \bibinfo{person}{Nestor
  Hernandez}, \bibinfo{person}{Ruben Recabarren},
  \bibinfo{person}{Syed~Ishtiaque Ahmed}, {and} \bibinfo{person}{Bogdan
  Carbunar}.} \bibinfo{year}{2019}\natexlab{}.
\newblock \showarticletitle{The Art and Craft of Fraudulent App Promotion in
  Google Play} {\em (\bibinfo{series}{CCS})}. \bibinfo{publisher}{ACM}.
\newblock


\bibitem[\protect\citeauthoryear{Redmiles, Chachra, and Waismeyer}{Redmiles
  et~al\mbox{.}}{2018}]%
        {redmiles2018examining}
\bibfield{author}{\bibinfo{person}{Elissa~M Redmiles}, \bibinfo{person}{Neha
  Chachra}, {and} \bibinfo{person}{Brian Waismeyer}.}
  \bibinfo{year}{2018}\natexlab{}.
\newblock \showarticletitle{Examining the demand for spam: Who clicks?}. In
  \bibinfo{booktitle}{{\em CHI}}.
\newblock


\bibitem[\protect\citeauthoryear{Reis, Melo, Garimella, and Benevenuto}{Reis
  et~al\mbox{.}}{2020}]%
        {reis2020can}
\bibfield{author}{\bibinfo{person}{Julio~CS Reis}, \bibinfo{person}{Philipe
  Melo}, \bibinfo{person}{Kiran Garimella}, {and}
  \bibinfo{person}{Fabr{\'\i}cio Benevenuto}.} \bibinfo{year}{2020}\natexlab{}.
\newblock \showarticletitle{Can WhatsApp benefit from debunked fact-checked
  stories to reduce misinformation?}
\newblock \bibinfo{journal}{{\em Harvard Misinformation Review\/}}
  (\bibinfo{year}{2020}).
\newblock


\bibitem[\protect\citeauthoryear{Resende et~al\mbox{.}}{Resende
  et~al\mbox{.}}{2019}]%
        {resende2019mis}
\bibfield{author}{\bibinfo{person}{Gustavo Resende} {et~al\mbox{.}}}
  \bibinfo{year}{2019}\natexlab{}.
\newblock \showarticletitle{(Mis) Information Dissemination in WhatsApp:
  Gathering, Analyzing and Countermeasures}. In \bibinfo{booktitle}{{\em Web
  Conference}}.
\newblock


\bibitem[\protect\citeauthoryear{Rintel, Mulholland, and Pittam}{Rintel
  et~al\mbox{.}}{2001}]%
        {rintel2001first}
\bibfield{author}{\bibinfo{person}{E~Sean Rintel}, \bibinfo{person}{Joan
  Mulholland}, {and} \bibinfo{person}{Jeffery Pittam}.}
  \bibinfo{year}{2001}\natexlab{}.
\newblock \showarticletitle{First things first: Internet relay chat openings}.
\newblock \bibinfo{journal}{{\em Journal of Computer-Mediated Communication\/}}
  (\bibinfo{year}{2001}).
\newblock


\bibitem[\protect\citeauthoryear{Saha, Mathew, Garimella, and Mukherjee}{Saha
  et~al\mbox{.}}{2021}]%
        {saha2021short}
\bibfield{author}{\bibinfo{person}{Punyajoy Saha}, \bibinfo{person}{Binny
  Mathew}, \bibinfo{person}{Kiran Garimella}, {and} \bibinfo{person}{Animesh
  Mukherjee}.} \bibinfo{year}{2021}\natexlab{}.
\newblock \showarticletitle{Short is the Road that Leads from Fear to Hate:
  Fear Speech in Indian WhatsApp Groups}. In \bibinfo{booktitle}{{\em Web
  Conference}}.
\newblock


\bibitem[\protect\citeauthoryear{Shin, Shim, Kim, Lee, Kang, and Hwang}{Shin
  et~al\mbox{.}}{2021}]%
        {shin2021twiti}
\bibfield{author}{\bibinfo{person}{Hyejin Shin}, \bibinfo{person}{WooChul
  Shim}, \bibinfo{person}{Saebom Kim}, \bibinfo{person}{Sol Lee},
  \bibinfo{person}{Yong~Goo Kang}, {and} \bibinfo{person}{Yong~Ho Hwang}.}
  \bibinfo{year}{2021}\natexlab{}.
\newblock \showarticletitle{\# Twiti: Social Listening for Threat
  Intelligence}. In \bibinfo{booktitle}{{\em Web Conference}}.
\newblock


\bibitem[\protect\citeauthoryear{Singer, Fl{\"o}ck, Meinhart, Zeitfogel, and
  Strohmaier}{Singer et~al\mbox{.}}{2014}]%
        {singer2014evolution}
\bibfield{author}{\bibinfo{person}{Philipp Singer}, \bibinfo{person}{Fabian
  Fl{\"o}ck}, \bibinfo{person}{Clemens Meinhart}, \bibinfo{person}{Elias
  Zeitfogel}, {and} \bibinfo{person}{Markus Strohmaier}.}
  \bibinfo{year}{2014}\natexlab{}.
\newblock \showarticletitle{Evolution of reddit: from the front page of the
  internet to a self-referential community?}. In \bibinfo{booktitle}{{\em Web
  Conference}}. ACM.
\newblock


\bibitem[\protect\citeauthoryear{Stringhini, Kruegel, and Vigna}{Stringhini
  et~al\mbox{.}}{2010}]%
        {stringhini2010detecting}
\bibfield{author}{\bibinfo{person}{Gianluca Stringhini},
  \bibinfo{person}{Christopher Kruegel}, {and} \bibinfo{person}{Giovanni
  Vigna}.} \bibinfo{year}{2010}\natexlab{}.
\newblock \showarticletitle{Detecting spammers on social networks}. In
  \bibinfo{booktitle}{{\em CCS}}.
\newblock


\bibitem[\protect\citeauthoryear{Thread-Reader}{Thread-Reader}{2020}]%
        {threadReader}
\bibfield{author}{\bibinfo{person}{Thread-Reader}.}
  \bibinfo{year}{2020}\natexlab{}.
\newblock \bibinfo{title}{Fake Flipkart website}.
\newblock   (\bibinfo{year}{2020}).
\newblock
\showURL{%
\url{bit.ly/fake-FK}}


\bibitem[\protect\citeauthoryear{Xiao, Freeman, and Hwa}{Xiao
  et~al\mbox{.}}{2015}]%
        {xiao2015detecting}
\bibfield{author}{\bibinfo{person}{Cao Xiao}, \bibinfo{person}{David~Mandell
  Freeman}, {and} \bibinfo{person}{Theodore Hwa}.}
  \bibinfo{year}{2015}\natexlab{}.
\newblock \showarticletitle{Detecting clusters of fake accounts in online
  social networks}. In \bibinfo{booktitle}{{\em Wksp.\ AI \& Security}}.
\newblock


\bibitem[\protect\citeauthoryear{Yardi, Romero, Schoenebeck,
  et~al\mbox{.}}{Yardi et~al\mbox{.}}{2010}]%
        {yardi2010detecting}
\bibfield{author}{\bibinfo{person}{Sarita Yardi}, \bibinfo{person}{Daniel
  Romero}, \bibinfo{person}{Grant Schoenebeck}, {et~al\mbox{.}}}
  \bibinfo{year}{2010}\natexlab{}.
\newblock \showarticletitle{Detecting spam in a twitter network}.
\newblock \bibinfo{journal}{{\em First Monday\/}} (\bibinfo{year}{2010}).
\newblock


\end{thebibliography}

\end{document}